\documentstyle[prb,aps,epsf]{revtex}

\begin{document}
\title{Surface Roughness and New Type of Size Effect in Quantized Films}
\author{A. E. Meyerovich, I. V. Ponomarev}
\address{Department of Physics, University of Rhode Island, 2 Lippitt Rd., Kingston\\
RI 02881-0817}
\date{\today}
\maketitle

\begin{abstract}
The effect of random surface roughness on quantum size effect in thin films
is discussed. The conductivity of quantized metal films is analyzed for
different types of experimentally identified correlation functions of
surface inhomogeneities including the Gaussian, exponential, power-law
correlators, and the correlators with a power law decay of the power density
spectral function. The dependence of the conductivity\ $\sigma $\ on the
film thickness $L$, correlation radius of inhomogeneities $R$, and the
fermion density is investigated. The goal is to help in extracting surface
parameters from transport measurements and to determine the importance of
the choice of the proper surface correlator for transport theory. A new type
of size effect is predicted for quantized films with large correlation
radius of random surface corrugation. The effect exists for inhomogeneities
with Gaussian and exponential power spectrum; if the decay of power spectrum
is slow, the films exhibit usual quantum size effect. The conductivity $%
\sigma $ exhibits well-pronounced oscillations as a function of the channel
width $L$ or the density of fermions, and large steps as a function of the
correlation radius $R.$ These oscillations and steps are explained and their
positions identified. This phenomenon, which is reminiscent of magnetic
breakthrough, can allow direct observation of the quantum size effect in
conductivity of nano-scale metal films. The only region with a nearly
universal behavior of transport is the region in which particle wavelength
is close to the correlation radius of surface inhomogeneities.
\end{abstract}

\pacs{72.10.Fk, 73.23.Ad,73.50.Bk}


\section{Introduction}

Progress in material technology, especially in nanofabrication, ultrathin
film manufacturing, ultraclean and high vacuum systems, {\it etc., }requires
better understanding of boundary scattering in physical processes. The
boundary effects should be an integral part of any study of quantum wires,
wells, and films. Boundary scattering is especially important for transport
in ultrathin and/or clean systems in which the particle mean free path is
comparable to the system size.

Below we consider the effect of random surface roughness on quantum
transport in quantized quasi-$2D$ systems such as, for example, ultrathin
metal films. The main issue is to find how sensitive is the transport along
such film to the statistical properties of random surface inhomogeneities
(thickness fluctuations). An important by-product of our systematic
comparison of different classes of random surface inhomogeneities is the
prediction of a new type of size effect in quantized films. This effect
manifests itself as large oscillations of conductivity $\sigma $ as a
function of the film thickness $L$. In contrast to the usual quantum size
effect (QSE), the peaks can be observed only at relatively large values of $%
L $. The distance between the peaks is large and is roughly proportional to $%
L^{2}$. The observation of this QSE opens a new experimental method of
identification of the type of surface roughness.

The choice of quasi-$2D$ systems is explained by a desire to avoid
divergence of surface fluctuations and strong localization effects which are
inherent to $1D$ systems and make a systematic quantitative study of the
effect of surface inhomogeneities on transport virtually impossible. In
contrast to $1D$ systems, the randomly fluctuating $2D$ surfaces are
practically stable while the localization length in systems with weak
surface roughness is exponentially large. (In general, the {\it transport}
problems are more interesting in systems with weak rather than with strong
roughness. Transport in systems with strong roughness is trivial: each wall
collision completely dephases the particles and the mean free path cannot
exceed the distance between the walls).

The prevalent way to characterize the random surface roughness and/or
thickness fluctuations is to use the correlation function of surface
inhomogeneities 
\begin{equation}
\zeta \left( {\bf s}\right) \equiv \zeta \left( \left| {\bf s}\right|
\right) =\left\langle \xi ({\bf s}_{1})\xi ({\bf s}_{1}+{\bf s}%
)\right\rangle \equiv A^{-1}\int \xi ({\bf s}_{1})\xi ({\bf s}_{1}+{\bf s})d%
{\bf s}_{1},  \label{aa1}
\end{equation}
where the vector ${\bf s}$ gives the $2D$ coordinates along the surface, $%
\xi ({\bf s})$\ describes the deviation of the position of the surface in
the point with $2D$ coordinates ${\bf s}$ from its average position, $%
\left\langle \xi ({\bf s})\right\rangle =0$, and $A$ is the averaging area.
Here it is\ assumed that the correlation properties of the surface do not
depend on direction. Two main characteristics of the surface correlation
functions $\zeta $ are the average amplitude (''height'') $\ell $ and
correlation radius (''size'') $R$ of surface inhomogeneities.

Any transport theory for systems with rough boundaries should provide the
explicit dependence of the particle mean free path (or the conductivity
along the walls) on the correlator of surface inhomogeneities $\zeta \left(
s\right) $. Without bulk scattering, the conductivity $\sigma $\ is
determined by the relation between three length scales: particle wavelength, 
$\Lambda $; the width of the channel, $L$; and the correlation radius of
inhomogeneities, $R$. If the roughness is weak, the fourth length parameter, 
$\ell $, enters conductivity as a coefficient, 
\begin{equation}
\sigma =\frac{2e^{2}}{\hbar }\frac{L^{2}}{\ell ^{2}}f\left( \Lambda
,L,R\right) .  \label{bb1}
\end{equation}
Note, that this $2D$ conductivity differs by a length unit from the usual $3D
$ conductivity and, as a result, has a dimensionality of conductance.

The form of the surface correlator $\zeta \left( s\right) $ can vary from
surface to surface. Most of the theoretical calculations assume that this
correlator is Gaussian. The numerical simulations, on the other hand, often
rely on various generators for random rough surfaces without paying much
attention to the correlation function of the generated inhomogeneities. Both
approaches are not satisfactory since the experiments on surface scattering
and diffraction patterns show that real surfaces exhibit surface correlators 
$\left( \text{\ref{aa1}}\right) $ of various forms\cite{q2,fer1}. Even one
and the same film can exhibit various correlation properties on different
stages of growth. As a result, the behavior of the functions $f\left(
\Lambda ,L,R\right) $ in Eq. $\left( \text{\ref{bb1}}\right) $, which
reflects the correlation properties of inhomogeneities, can vary from
surface to surface even when the main correlation parameters $\ell $ and $R$
remain the same.

The correlation functions $\left( \text{\ref{aa1}}\right) $ are
characterized by different long-range behavior that can be reliably
identified in various surface diffraction and scattering experiments. What
we would like to know is how sensitive is the {\it particle} {\it transport}
to the form of the surface correlator. In contrast to surface diffraction
and scattering data with angular and/or wavelength scanning, the transport
coefficients are integral parameters that include angular and wavelength
averaging. This leaves the question of how sensitive is the conductivity to
the shape of the surface correlator wide open. In addition, we are asking a
question whether it is possible to identify the type of surface
inhomogeneities from transport experiments in ultrathin films or multilayer
systems without beforehand information on the form of the surface
correlator. The interrelated question is, of course, to what extent one
should pay attention to the details of the correlator of surface
inhomogeneities in analytical or numerical transport calculations for
particles with large mean free paths. The former issue has already been
raised in Refs.\cite{fish2,pal1} for a small set of surface correlators on
the basis of the Born approximation for wall scattering. Below we present a
systematic study which is based on a more general transport formalism and
involves a variety of classes of surface correlators.

In short, we want to compare functions $f\left( \Lambda ,L,R\right) $ in Eq.$%
\left( \text{\ref{bb1}}\right) $ calculated for various types of the
correlation functions $\zeta \left( s\right) $ in a wide range of
parameters. We start from degenerate ballistic fermions in quantized metal
films. The choice is not arbitrary: transport in such systems involves the
minimal degree of averaging (integration) and can be the most sensitive to
the long-range properties of the surface correlators $\left( \text{\ref{aa1}}%
\right) $.

Quantum size effect (QSE) in metal films is a subject of intensive
experimental study. Recent QSE experiments with quantized metal films
include conductivity \cite{conduct1}, spectroscopy \cite{spect1},
susceptibility \cite{susc1}, and STM \cite{stm1} measurements. One of the
signature features of QSE in metals is a pronounced saw-like dependence of
conductivity on, for example, film thickness, $\sigma \left( L\right) $.
This dependence was predicted for both bulk \cite{rr6} and surface \cite{r4}
scattering. Experimentally, QSE in conductivity was studied for metals in
Refs.\cite{conduct1,qse2} (for earlier results see references therein).
However, experiments on QSE in metals have to overcome a difficulty which
one does not encounter in semiconductors. The period of the QSE oscillations
in the dependence $\sigma \left( L\right) $ is usually small, almost atomic, 
$1/p_{F}$ (below, except for final results, $\hbar =1$). For this reason
typical experimental object are lead or semimetal films such as bismuth.
Below we predict a new type of QSE with large-period oscillations of $\sigma
\left( L\right) $ at relatively large values of $p_{F}L$ that could lead to
observation of QSE in a wider group of metals. Large-period QSE oscillations
have already been observed (see the second Ref. \cite{conduct1}); however,
sketchy experimental details do not allow one to identify reliably this
observation as the new type of QSE predicted below. Our results can also
resolve the long-standing controversy on the influence of the structure of
the nanoscale film on its resistivity \cite{qse2}.

Recently, we developed a transparent semi-analytical formalism for transport
in systems with rough boundaries that allows simple uniform calculations in
a wide range of parameters and for various types of roughness with and
without bulk scattering \cite{arm1,arm2,arm3}. This formalism unites
approaches by Tesanovic {\it et al } \cite{r3}, Fishman and Calecki \cite
{qq17}, Kawabata \cite{kaw1}, Meyerovich and S. Stepaniants \cite{r2}, and
Makarov {\it et al}\cite{mak1} (for a brief comparison between different
theoretical approaches see Refs.\cite{arm2,arm4}). Below we apply this
formalism with an explicit purpose of studying the dependence of the
transport coefficients on the shape of the correlation function of surface
inhomogeneities. The well-defined limits of applicability of our approach to
metal and semiconductor films are discussed in detail in Refs.\cite
{arm2,arm3}.

Since the $2D$ mobility of particles is described by essentially the same
equations as the exponent in the expression for the localization length in
films, our study provides the dependence of the localization length on the
type of the correlation function of random surface inhomogeneities.

The paper has the following structure. In the next Section we introduce
various types of surface correlation functions. Section III briefly
describes the transport equation used for conductivity (mobility)
calculations in QSE conditions. The results of transport calculations for
different types of correlators are given in Section IV. Conclusions and
experimental implications are discussed in Section V. Appendix A contains
useful analytical expressions for the power density spectral functions of
inhomogeneities responsible for the behavior of scattering probabilities for
different types of correlators. Appendix B deals with the positions of new
type of QSE peaks.

\section{Correlation function of surface inhomogeneities}

We consider an infinite $2D$ channel (or film) of the average thickness $L$
with random rough boundaries 
\begin{equation}
x=L/2-\xi _{1}(y,z),\ x=-L/2+\xi _{2}(y,z).  \label{a0}
\end{equation}
(the walls are assumed hard with infinite potential). The inhomogeneities
are small, $\xi _{1,2}\left( y,z\right) \ll L$, and random with zero
average, $\left\langle \xi _{1}\right\rangle =\left\langle \xi
_{2}\right\rangle =0$. Their correlation function $\zeta _{ik}\left( {\bf s}%
\right) $ and its Fourier image $\zeta _{ik}\left( {\bf q}\right) $, which
is often called the power spectral density function or power spectrum, are
defined as 
\begin{eqnarray}
\zeta _{ik}\left( \left| {\bf s}\right| \right)  &=&\left\langle \xi _{i}(%
{\bf s}_{1})\xi _{k}({\bf s}_{1}+{\bf s})\right\rangle \equiv A^{-1}\int \xi
_{i}({\bf s}_{1})\xi _{k}({\bf s}_{1}+{\bf s})d{\bf s}_{1},  \label{a1} \\
\zeta _{ik}\left( \left| {\bf q}\right| \right)  &=&\int d^{2}s\ e^{i{\bf %
q\cdot s}}\zeta _{ik}\left( \left| {\bf s}\right| \right) =2\pi
\int_{0}^{\infty }\zeta _{ik}\left( s\right) J_{0}\left( qs\right) sds 
\nonumber
\end{eqnarray}
where ${\bf s=}\left( y,z\right) $ and ${\bf q=}\left( q_{y},q_{z}\right) $
are the $2D$ vectors. In homogeneous systems, the correlation function
depends only on the distance between points $\left| {\bf s}_{1}-{\bf s}%
_{2}\right| $ and not on coordinates themselves. The correlation functions $%
\zeta _{11}$ and $\zeta _{22}$ describe intrawall correlations of
inhomogeneities, and $\zeta _{12}=\zeta _{21}$ are the interwall
correlations. Usually, but not always, the inhomogeneities on different
walls are not correlated with each other, $\zeta _{12}=0$. Thus, everywhere,
except for Section IVE, it is assumed that $\zeta _{12}=0$. To avoid
parameter clutter, we also assume that the correlation parameters are the
same on both walls, $\zeta _{11}=\zeta _{22}=\zeta $. Then the effective
correlator contains $2\zeta \left( s\right) $ with $\zeta \left( s\right) $
given by equations below.

Surface inhomogeneities exhibit a variety of types of the correlation
functions \cite{q2,fer1}. To have a meaningful comparison, we consider the
correlation functions that involve only two characteristic parameters,
namely, the amplitude (average height) $\ell $ and the correlation radius
(average size) $R$ of surface inhomogeneities.

The most commonly used in theoretical applications is the Gaussian
correlation function, 
\begin{equation}
\zeta \left( s\right) =\ell ^{2}\exp \left( -s^{2}/2R^{2}\right) ,\ \zeta
\left( q\right) =2\pi \ell ^{2}R^{2}\exp \left( -q^{2}R^{2}/2\right)
\label{a2}
\end{equation}
including its limit for small correlation radius $R\rightarrow 0$, {\it i.e.,%
} the $\delta $-type correlations, 
\begin{equation}
\zeta \left( s\right) =\ell ^{2}R^{2}\delta \left( s\right) /s,\ \zeta
\left( q\right) =2\pi \ell ^{2}R^{2}.  \label{a3}
\end{equation}

Sometimes, a better fit to experimental data on surface scattering is
provided by the use of the exponential correlation function 
\begin{equation}
\zeta \left( s\right) =\ell ^{2}\exp \left( -s/R\right) ,\ \zeta \left(
q\right) =\frac{2\pi \ell ^{2}R^{2}}{\left( 1+q^{2}R^{2}\right) ^{3/2}},
\label{e2}
\end{equation}
or by the even more long-range, power-law correlators 
\begin{equation}
\zeta \left( s\right) =\frac{2\mu \ell ^{2}}{\left( 1+s^{2}/R^{2}\right)
^{1+\mu }},\ \zeta \left( q\right) =2\pi \ell ^{2}R^{2\,}\frac{\left(
qR\,\right) ^{\mu }}{2^{\mu -1}\Gamma \left( \mu \right) }K_{\mu }\left(
qR\right)  \label{ee2}
\end{equation}
with different values of the parameter $\mu $. The most commonly used are
the Staras function with $\mu =1$ and the correlator with $\mu =1/2$ which
has the exponential power spectrum $\zeta \left( q\right) $, 
\begin{equation}
\zeta \left( q\right) =2\pi \ell ^{2}R^{2}\exp \left( -qR\right) \text{.}
\label{ee22}
\end{equation}

The use of the Lorentzian correlator, which differs from the definition $%
\left( \text{\ref{ee2}}\right) $ at $\mu \rightarrow 0$ by the factor $\mu $
in the numerator$,$ 
\begin{equation}
\zeta \left( s\right) =\frac{2\ell ^{2}}{1+s^{2}/R^{2}},\ \zeta \left(
q\right) =2\pi \ell ^{2}R^{2\,}K_{0}\left( qR\right) ,  \label{eee2}
\end{equation}
deserves a special comment. This correlator is often considered as
''unphysical''. Its Fourier image\ $\left( \text{\ref{eee2}}\right) $
contains function $K_{0}\left( qR\right) $\ that diverges logarithmically at
long wavelengths $q\rightarrow 0$. The issue to what extent the correlators
are ''physical'' and can be reproduced experimentally is irrelevant in our
context. For us, the fact that the Lorentzian correlator is sometimes used
in calculations  is sufficient enough to consider this correlator in the
paper. To deal with the divergency, one can truncate the Lorentzian
correlator at large distances (the common practice is to make a cut-off at
the distances about 0.1 of the system length \cite{q2}). Another option is
to use the generalized power-law correlator $\left( \text{\ref{ee2}}\right) $
with small $\mu $ instead of the Lorentzian $\left( \text{\ref{eee2}}\right) 
$. In order not to introduce additional parameters, we use the untruncated
equation $\left( \text{\ref{eee2}}\right) $. Even though the divergence of $%
K_{0}\left( qR\rightarrow 0\right) $ does not lead to any singularities in
transport coefficients, the transport coefficients for Lorentzian surfaces
(see below) often behave qualitatively different from systems with other
types of random inhomogeneities, even from the systems $\left( \text{\ref
{ee2}}\right) $ with small $\mu $. (Sometimes, the divergence of the power
spectrum $\zeta \left( q\right) $ is associated with the fractal nature of
the surface \cite{q2}; to what extent our transport formalism can be used
for films with fractal surfaces is an open question).

The last class of correlation functions covers the power-law correlators in
momentum space, 
\begin{equation}
\zeta \left( q\right) =\frac{2\pi \ell ^{2}R^{2}}{\left( 1+q^{2}R^{2}\right)
^{1+\lambda }},\ \zeta \left( s\right) =\ell ^{2}\frac{\left( s/R\,\right)
^{\lambda }}{2^{\lambda }\Gamma \left( 1+\lambda \right) }K_{\lambda }\left(
s/R\right) .  \label{ee3}
\end{equation}
The correlators from this group include the Lorentzian in momentum space $%
\lambda =0$ that was observed in Ref.\cite{fer1} (see also Ref{\it .}\cite
{pal1}) and the exponential correlator $\left( \text{\ref{e2}}\right) $ at $%
\lambda =1/2$.

The constants in all these correlators are chosen in such a way that the
value of $\zeta \left( q=0\right) =2\pi \ell ^{2}R^{2}$ is the same. This
provides a reasonable basis of comparison for transport coefficients in
films with all these different types of random surfaces. Indeed, the
scattering cross-section for $q\rightarrow 0$ does not depend on the details
of short- and mid-range structure of surface inhomogeneities. Therefore, at
Fermi momenta $q_{F}\rightarrow 0$ (more precisely, at $q_{F}R\ll 1$), the
transport coefficients should be the same for all random surfaces. (The only
exception is the Lorentzian $\left( \text{\ref{eee2}}\right) $ for which $%
\zeta \left( q\right) $ diverges at small $q$).

In what follows we compare the transport properties of the films $\left( 
\text{\ref{a2}}\right) $ - $\left( \text{\ref{ee3}}\right) $\ in various
ranges of the film thickness $L$, correlation radius $R$, and the particle
wavelength $\Lambda _{F}=1/q_{F}$ (or the $2D$ particle density $N$).

\section{Transport equation for ballistic degenerate fermions in quantized
films}

QSE is caused by quantization of motion in the direction perpendicular to
the film, $p_{x}\rightarrow \pi j/L$, and leads to a split of the energy
spectrum $\epsilon \left( {\bf p}\right) $ into a set of minibands, $%
\epsilon \left( p_{x},{\bf q}\right) \rightarrow \epsilon \left( \pi j/L,%
{\bf q}\right) =\epsilon _{j}\left( {\bf q}\right) $. For simplicity, we
consider spherical Fermi surfaces $\epsilon _{j}\left( {\bf q}\right)
=\epsilon _{F}$, 
\begin{equation}
\epsilon _{j}\left( {\bf q}\right) =\frac{1}{2m}\left[ \left( \pi j/L\right)
^{2}+q_{j}^{2}\right] ,\ q_{j}\equiv q_{Fj}=\left[ 2m\epsilon _{F}-\left(
\pi j/L\right) ^{2}\right] ^{1/2},  \label{t1}
\end{equation}
where $q_{j}$ is the Fermi momentum for the miniband $j$. One can introduce
the overall Fermi momentum as 
\begin{equation}
q_{F}=1/\Lambda _{F}=\left( 2m\epsilon _{F}\right) ^{1/2}.  \label{c1}
\end{equation}
The relationship between this Fermi momentum $q_{F}$ and the $2D$ density of
fermions $N_{2}$ in quantized films is somewhat cumbersome \cite{arm1}: 
\begin{equation}
N=\sum N_{j}=\left( S/2\pi \right) \left[ q_{F}^{2}-\left( \pi /L\right)
^{2}\left( S+1\right) \left( 2S+1\right) /6\right] ,  \label{cc1}
\end{equation}
where $S$ is the number of the occupied minibands, 
\begin{equation}
S={\rm Int}\left[ q_{F}L/\pi \right]  \label{ee12}
\end{equation}
If the density of fermions is the same as in the bulk, then $N_{2}=n_{3}L$
where $n_{3}$ is the usual bulk density. Even in this case, the number of
the occupied minibands $S$, according to Eqs. $\left( \text{\ref{cc1}}%
\right) ,\left( \text{\ref{ee12}}\right) $,$\ $is a complicated function of $%
L$. Asymptotically, at large $S$ 
\begin{equation}
S={\rm Int}\left[ \left( 3N_{2}L^{2}/\pi \right) ^{1/3}\right] .
\label{eee12}
\end{equation}

According to Refs. \cite{arm1,arm2}, scattering by random surface
inhomogeneities results in intra- and interband transitions $\epsilon
_{j}\left( {\bf q}\right) \rightarrow \epsilon _{j^{\prime }}\left( {\bf q}%
^{\prime }\right) $ with transition probabilities $W_{jj^{\prime }}\left( 
{\bf q},{\bf q}^{\prime }\right) $ that are expressed explicitly via the
surface correlation function $\zeta \left( \left| {\bf q-q}^{\prime }\right|
\right) $: 
\begin{equation}
W_{jj^{\prime }}\left( {\bf q},{\bf q}^{\prime }\right) =\frac{\hbar }{%
m^{2}L^{2}}\left[ \zeta _{11}+\zeta _{22}+2\zeta _{12}\left( -1\right)
^{j+j^{\prime }}\right] \left( \frac{\pi j}{L}\right) ^{2}\left( \frac{\pi
j^{\prime }}{L}\right) ^{2}.  \label{rrr3}
\end{equation}
The generalization to other, more complicated energy spectra is
straightforward \cite{arm2}.

The transport equation for the distribution functions $n_{j}\left( {\bf q}%
\right) ,$ 
\begin{equation}
\frac{dn_{j}}{dt}=2\pi A\sum_{j^{\prime }}\int W_{jj^{\prime }}\left[
n_{j^{\prime }}-n_{j}\right] \delta \left( \epsilon _{j{\bf q}}-\epsilon
_{j^{\prime }{\bf q}^{\prime }}\right) \frac{d^{2}q^{\prime }}{\left( 2\pi
\right) ^{2}},  \label{aa3}
\end{equation}
reduces, after standard transformations, to a set of linear equations 
\begin{eqnarray}
q_{j}/m &=&-\sum_{j^{\prime }}\nu _{j^{\prime }}\left( q_{j^{\prime
}}\right) /\tau _{jj^{\prime }},  \label{ee6} \\
\frac{2}{\tau _{jj^{\prime }}} &=&m\sum_{j^{\prime \prime }}\left[ \delta
_{jj^{\prime }}\,W_{jj^{\prime \prime }}^{\left( 0\right) }-\delta
_{j^{\prime }j^{\prime \prime }}\,W_{jj^{\prime }}^{\left( 1\right) }\right]
\nonumber
\end{eqnarray}
where $n_{j}^{\left( 1\right) }=\nu _{j}\delta \left( \epsilon -\epsilon
_{F}\right) eE$ is the first angular harmonic of the distribution function $%
n_{j}\left( {\bf q}\right) $ at $q=q_{j}$, and $W_{jj^{\prime }}^{\left(
0,1\right) }\left( q_{j},q_{j^{\prime }}\right) $ are the zeroth and first
harmonics of $W\left( {\bf q}_{j}{\bf -q}_{j^{\prime }}\right) $\ over the
angle $\widehat{{\bf q}_{j}{\bf q}}_{j^{\prime }}$. For some of the
correlation function from the previous Section the angular harmonics can be
calculated analytically (see Appendix A). For others, this calculation is
performed numerically.

The solution of Eqs. $\left( \text{\ref{ee6}}\right) $ provides the
conductivity of the film: 
\begin{equation}
\sigma =-\frac{e^{2}}{3\hbar ^{2}}\sum_{j}\nu _{j}\left( q_{j}\right) q_{j}.
\label{ee7}
\end{equation}

Equations $\left( \text{\ref{ee6}}\right) $ have simple analytical solution
when the matrix $\tau _{jj^{\prime }}^{-1}$ can be approximated by a
diagonal matrix, $\tau _{jj^{\prime }}^{-1}\approx \delta _{jj^{\prime
}}/\tau _{j},$%
\begin{equation}
\sigma =\frac{e^{2}}{3\hbar ^{2}m}\sum_{j}q_{j}^{2}\tau _{j}  \label{en7}
\end{equation}
This happens when the matrix $W_{jj^{\prime }}^{\left( 1\right) }$ is almost
or exactly diagonal, $W_{jj^{\prime }}^{\left( 1\right) }\simeq
W_{j}^{\left( 1\right) }\delta _{jj^{\prime }}$ and 
\begin{equation}
2/m\tau _{j}=\sum_{j^{\prime }}W_{jj^{\prime }}^{\left( 0\right)
}-W_{j}^{\left( 1\right) }.  \label{eee7}
\end{equation}
Then the conductivity $\left( \text{\ref{en7}}\right) $ is equal to 
\begin{equation}
\sigma =\frac{e^{2}}{3\hbar ^{2}m}\sum_{j}\tau _{j}q_{j}^{2}=\frac{2e^{2}}{%
3\hbar ^{2}m^{2}}\sum_{j}\frac{q_{j}^{2}}{\sum_{j^{\prime }}W_{jj^{\prime
}}^{\left( 0\right) }-W_{j}^{\left( 1\right) }}.  \label{ee8}
\end{equation}

Such a diagonalization occurs in three physical situations. The simplest one
is the one when only one miniband is occupied and 
\begin{equation}
\sigma =\frac{e^{2}}{3\hbar ^{2}m}\tau _{1}q_{1}^{2}=\frac{2e^{2}q_{1}^{2}}{%
3\hbar ^{2}m^{2}}\frac{1}{W_{11}^{\left( 0\right) }-W_{11}^{\left( 1\right) }%
}.  \label{ee9}
\end{equation}

The second case is the case of systems with large correlation length $R\gg L$%
. In such systems the intraband scattering is much stronger than the
interband one and the off-diagonal matrix elements $W_{jj^{\prime }}$\ are
small in comparison with the diagonal ones (see Appendix A). Then both
matrices $W_{jj^{\prime }}^{\left( 0,1\right) }$ are almost diagonal,

\begin{equation}
W_{jj^{\prime }}^{\left( 0,1\right) }\simeq W_{j}^{\left( 0,1\right) }\delta
_{jj^{\prime }},  \label{eee9}
\end{equation}
and the expression for the conductivity $\left( \text{\ref{ee8}}\right) $
reads 
\begin{equation}
\sigma \simeq \frac{2e^{2}}{3\hbar ^{2}m^{2}}\sum_{j}\frac{q_{j}^{2}}{%
W_{j}^{\left( 0\right) }-W_{j}^{\left( 1\right) }}.  \label{ee10}
\end{equation}
Such diagonalization of the matrices $W_{jj^{\prime }}^{\left( 0,1\right) }$ 
$\left( \text{\ref{eee9}}\right) $\ at $R\gg L$ can often be an
oversimplification (see Section IV).

The third situation with diagonal $\tau _{jj^{\prime }}^{-1}$ is the case of
small $qR$. In this limit, the correlation function is a constant with the
zero first harmonic, 
\[
W_{jj}^{\left( 0\right) }=2W\left( qR\rightarrow 0\right) ,\ W_{jj}^{\left(
1\right) }=0 
\]
According to Eq.$\left( \text{\ref{rrr3}}\right) ,$ 
\begin{equation}
W_{jj^{\prime }}\left( 0\right) =\frac{2\hbar }{m^{2}L^{2}}\zeta \left(
0\right) \left( \frac{\pi j}{L}\right) ^{2}\left( \frac{\pi j^{\prime }}{L}%
\right) ^{2}.  \label{eee10}
\end{equation}
and 
\begin{equation}
\sigma =\frac{2e^{2}}{\hbar }\frac{\left( L/\pi \right) ^{4}}{2S\left(
S+1\right) \left( 2S+1\right) \zeta \left( 0\right) }\sum_{j}\left( \frac{%
Lq_{j}}{\hbar j}\right) ^{2}.  \label{ee11}
\end{equation}

Note, that all our surface correlators $\zeta \left( s\right) $ are
introduced in such a way that in the longwave limit $\zeta \left(
q\rightarrow 0\right) $ they are, except for the Lorentzian $\left( \text{%
\ref{eee2}}\right) $, equal to each other, $\zeta \left( 0\right) =2\pi \ell
^{2}R^{2}.$ This means that in this limit the conductivities $\left( \text{%
\ref{ee11}}\right) $ are the same irrespective of the shape of the
correlator, 
\begin{equation}
\sigma =\frac{2e^{2}}{\hbar }\frac{1}{4\pi }\frac{\left( L^{2}/\pi ^{2}\ell
R\right) ^{2}}{S\left( S+1\right) \left( 2S+1\right) }\sum_{j}\left( \frac{%
Lq_{j}}{\hbar j}\right) ^{2}.  \label{ee13}
\end{equation}
({\it cf.} Ref.\cite{qq17}).\ 

In all other situations Eqs.$\left( \text{\ref{ee6}}\right) $ are not
diagonal and should be solved numerically.

The results for conductivity (mobility) also provide the exponent in the
expression for the localization length ${\cal R}$ that describes
localization caused by particle scattering by random wall inhomogeneities 
\cite{arm2}: 
\begin{equation}
{\cal R}={\cal L}\exp \left[ \pi mSD/\hbar \right]  \label{loc1}
\end{equation}
where ${\cal L}$ is the mean free path and the diffusion coefficient $D$ is
proportional to the conductivity $\sigma $.

\section{Results and discussion}

\subsection{General comments}

As it is mentioned in Introduction, the $2D$ conductivity $\sigma $\ of the
film has the dimensionality of conductance and is described by a
dimensionless function $f\ $in Eq. $\left( \text{\ref{bb1}}\right) .$\ This
function, in turn, depends on the relation between three length scales -
particle Fermi wavelength $\Lambda _{F}=1/q_{F},$ the width of the channel $%
L $, and the correlation radius of the surface inhomogeneities $R$. The
fourth length parameter, $\ell $, is perturbative and enters conductivity as
a coefficient, 
\begin{equation}
\sigma =\frac{2e^{2}}{\hbar }\frac{L^{2}}{\ell ^{2}}f\left( \Lambda
_{F},L,R\right) .  \label{c2}
\end{equation}
Note, that we consider only the contribution from surface roughness and
disregard bulk scattering. As a result, the conductivity $\left( \text{\ref
{c2}}\right) $ diverges in the limit of vanishing inhomogeneities $\ell
\rightarrow 0$\ or $R\rightarrow \infty $. The proper account of bulk
scattering \cite{arm3} eliminates this divergence.

The dimensionless function $f\left( \Lambda _{F},L,R\right) $ depends only
on the ratio of these three lengths. Of three ratios $z=L/\Lambda
_{F}=q_{F}L $, $x=R/\Lambda _{F}=q_{F}R$, and $y=R/L=x/z$ only two are
independent, $x=yz $. Which two of these ratios should be used as
independent dimensionless variables depends on whether one wants to display
the dependence of $\sigma $ on $\Lambda _{F}$, $L$, or $R$. The study of the
dependence of the conductivity on film thickness, $\sigma \left( L\right) $,
should be performed at constant $\Lambda _{F}$\ and $R$. This means that $%
\sigma \left( L\right) $ is best displayed by the function $f_{L}\left(
z,x\right) , $ 
\begin{equation}
\sigma \left( L\right) =\frac{2e^{2}}{\hbar }\frac{R^{2}}{\ell ^{2}}%
f_{L}\left( z,x=const\right) ,  \label{c3}
\end{equation}
for various values of $x=R/\Lambda _{F}$.

Plots of the function $f_{R}\left( y\right) $ at constant values of $%
z=q_{F}L,$%
\begin{equation}
\sigma \left( R\right) =\frac{2e^{2}}{\hbar }\frac{L^{2}}{\ell ^{2}}%
f_{R}\left( y,z=const\right) ,  \label{c4}
\end{equation}
reflect the dependence $\sigma \left( R\right) $. Similarly, plots of the
function $f_{N}\left( z\right) $ at constant $y=R/L,$%
\begin{equation}
\sigma \left( q_{F}\right) =\frac{2e^{2}}{\hbar }\frac{L^{2}}{\ell ^{2}}%
f_{N}\left( z,y=const\right) ,  \label{c5}
\end{equation}
characterizes the dependence of conductivity on density of particles $N$ or
the Fermi momentum $q_{F}$.

Below we compare these dimensionless functions, $f_{L}\left( z\right) ,$ $%
f_{R}\left( y\right) ,$ and $f_{N}\left( z\right) $ for various types of
correlation functions in wide ranges of parameters. Needless to say, the
results at $x\rightarrow 0$ should coincide for all types of correlators
except, maybe, for the Lorentzian, since, by design, all the correlation
functions are the same in this limit [see Eq.\ $\left( \text{\ref{ee13}}%
\right) $].

Curves in all figures below are labeled in a uniform way by the type of
surface correlator used in calculations. Curves $G$ correspond to Gaussian
inhomogeneities $\left( \text{\ref{a2}}\right) $, Curves $L$ describe the
surfaces with Lorentzian correlations $\left( \text{\ref{eee2}}\right) $,
Curves $\mu _{1}$, $\mu _{5}$, and $\mu _{9}$ give the results for the
correlators $\left( \text{\ref{ee2}}\right) $\ with $\mu =0.1;0.5;09$, and
Curves $\lambda _{0}$, $\lambda _{5}$, and $\lambda _{9}$ correspond to Eq. $%
\left( \text{\ref{ee3}}\right) $\ with $\lambda =0;0.5;0.9$. Note, that
correlator $\mu _{5}$ has the exponential power spectrum $\left( \text{\ref
{ee22}}\right) $ and that correlator $\lambda _{5}$ is actually the
exponential correlator $\left( \text{\ref{e2}}\right) $.

\subsection{Dependence on the film thickness}

Figures 1 - 2 for the function $f_{L}\left( z,x=const\right) $, Eq. $\left( 
\text{\ref{c3}}\right) $, show the dependence of the conductivity $\sigma
\left( L\right) $ for two different values of $R/\Lambda _{F}$, $x=1;10$,
for various types of the correlation functions. The labeling of the curves $%
G,L,\mu _{1},\mu _{5},\mu _{9},\lambda _{0},\lambda _{5},\lambda _{9}$ is
explained in the end of previous subsection. The main feature of the curves,
namely, their saw-like character, is well known. The sharp drops occur when
the number of the occupied minibands, Eq$.$ $\left( \text{\ref{ee12}}\right) 
$ changes by 1, {\it i.e., }in the points $z=L/\Lambda _{F}=k\pi $ with
integer $k.$ The only unexpected feature is a ''wrong'' periodicity of the
initial part of the Gaussian curve $G$ at small values of $z$ for $x=10$
(see insert in Figure 2). This feature will be explained later. The
Lorentzian curve $L$ is different from others: at $x=10$ the curve has
already lost its QSE structure. 
\begin{figure}[tbp]
\centerline{\epsfxsize=3.4in\epsfbox{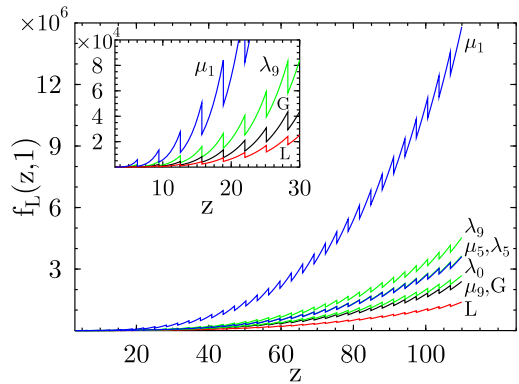}}
\caption{Function $f_{L}\left( z,x=const\right) $, Eq.$\left( \text{\ref{c3}}%
\right) $, at $x=R/\Lambda _{F}=1$ for various correlation functions. The
labeling of the curves is explained in the end of Sec. IVA. Curve $G$:
Gaussian correlator $\left( \text{\ref{a2}}\right) $;\ curves $\protect\mu %
_{1},\protect\mu _{5},\protect\mu _{9}$: power-law correlators $\left( \text{%
\ref{ee2}}\right) $ with $\protect\mu =0.1;0.5,0.9$; curve $L$: Lorentzian
correlator; curves $\protect\lambda _{0},\protect\lambda _{5},\protect\lambda
_{9}$: power-law correlators in momentum space $\left( \text{\ref{ee3}}%
\right) $\ with $\protect\lambda =0;0.5;0.9$ ($\protect\lambda =0.5$
corresponds to the exponential correlator in the coordinate space $\left( 
\text{\ref{e2}}\right) $). The sharp drops occur when the number of the
occupied minibands $S,$ Eq.$\left( \text{\ref{ee12}}\right) ,$ changes by 1, 
{\it i.e., }in the points $z=L/\Lambda _{F}=k\protect\pi $ with integer $k$. 
}
\label{fig1}
\end{figure}
At these, relatively small values of $x,$ the curves for all types of
correlators have roughly the same shape though the exact values of the
conductivity are different. (The curves $\mu _{5}$ and $\lambda _{5}$ are
indistinguishable in both Figures 1 and 2, and curves $G$ and $\mu _{9}$ are
indistinguishable in Figure 1). To underscore this point, in Figures 3 and 4
we plotted instead of the curves $f_{L}\left( z\right) $ the normalized
curves $f_{L}\left( z\right) /f_{L}\left( z=z_{\max }\right) $ with the
normalization coefficients ensuring that the values of the normalized
conductivity are equal to 1 at the highest values of $z$ in the plot.
Strikingly, for $x=1$ (Figure 3) all the normalized curves with these 8
correlation functions lie within one bold line and are {\it all}
indistinguishable with this resolution. For larger $x$, the difference is
still insignificant: all the curves are compressed between curves $G$ and $L$%
. The only anomaly is the loss of QSE structure by curve $L$. 
\begin{figure}[tbp]
\centerline{\epsfxsize=3.4in\epsfbox{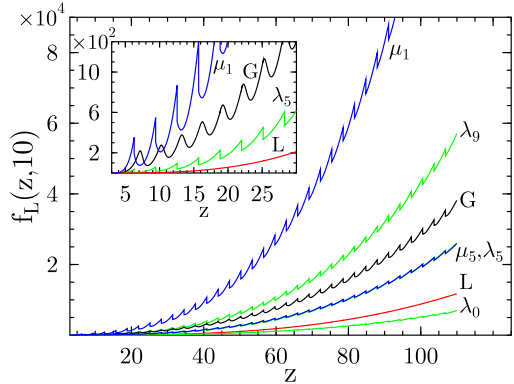}}
\caption{The same as in Figure 1 for $x=10$. The labeling of the curves is
explained in the end of Sec. IVA. }
\label{fig2}
\end{figure}
\begin{figure}[tbp]
\centerline{\epsfxsize=3.4in\epsfbox{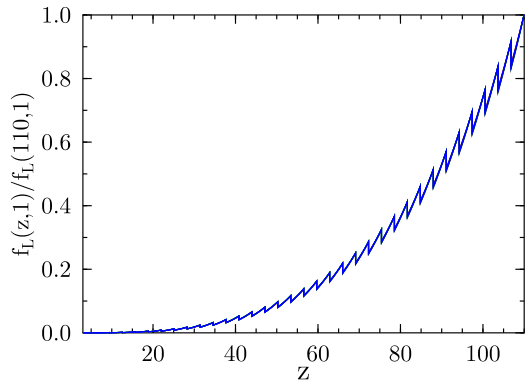}}
\caption{The same 8 functions $f_{L}\left( z,x=1\right) $ as in Figure 1
normalized by their value at $z=110$, $f_{L}\left( z\right) /f_{L}\left(
110\right) $. All 8 normalized curves are indistinguishable. The
normalization coefficients are: curve $G$ - $f_{L}\left( 110\right)
=2.4\cdot 10^{6}$; curve $L$ - $f_{L}\left( 110\right) =1.39\cdot 10^{6}$;
curve $\protect\mu _{1}$ - $f_{L}\left( 110\right) =1.48\cdot 10^{7}$; curve 
$\protect\mu _{5}$ - $f_{L}\left( 110\right) =3.61\cdot 10^{6}$; curve $%
\protect\mu _{9}$ - $f_{L}\left( 110\right) =2.42\cdot 10^{6}$; curve $%
\protect\lambda _{0}$ - $f_{L}\left( 110\right) =2.69\cdot 10^{6}$; curve $%
\protect\lambda _{5}$ - $f_{L}\left( 110\right) =3.65\cdot 10^{6}$; curve $%
\protect\lambda _{9}$ - $f_{L}\left( 110\right) =4.54\cdot 10^{6}$. }
\label{fig3}
\end{figure}

The main conclusion here is that the {\it shape} of the dependence $\sigma
\left( L\right) $ at constant $R$ and $q_{F}$ is not sensitive to and cannot
provide any information on the type of the correlator at not very large
values of $R/\Lambda _{F}$. Since $\ell $ is unknown and enters the
conductivity as a coefficient, the absolute values of $\sigma \left(
L\right) $ cannot serve as a clue either: experimental data on $\sigma
\left( L\right) $ at moderate $R/\Lambda _{F}$ can be fitted by any type of
the correlator by a choice of $\ell $. In this case, it is impossible to
make any conclusion on the type of correlation function from transport
measurements and it does not matter what correlator to use in theoretical
calculations. Meaningful analysis requires some{\it \ }beforehand{\it \ }%
information on the correlation parameters. The only correlator that can be
identified is the Lorentzian; however, this type of correlations is the
least probable and might be ''unphysical''.

The situation changes dramatically at higher $x=R/\Lambda _{F}$ as is shown
in Figures 5 [function $f_{L}\left( z,x=400\right) $] and 6 [normalized
function $f_{L}\left( z,x=400\right) /f_{L}\left( z=z_{\max },x=400\right) $%
] for the same 8 correlators (the labeling of the curves is explained in the
end of Sec. IVA).
\begin{figure}[tbp]
\centerline{\epsfxsize=3.4in\epsfbox{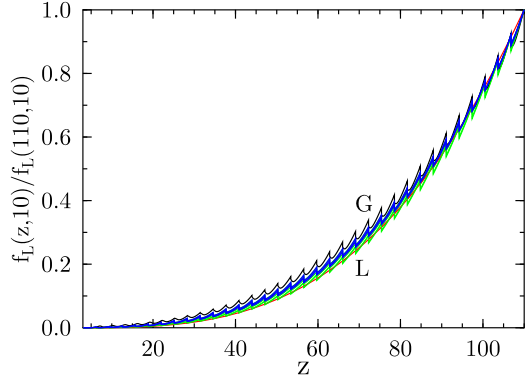}}
\caption{The same 8 functions $f_{L}\left( z,x=10\right) $ as in Figure 2
normalized by their value at $z=110$, $f_{L}\left( z\right) /f_{L}\left(
110\right) $. All 8 curves lie between normalized curves $G$ and $L$ and are
barely distinguishable. The normalization coefficients are: curve $G$ - $%
f_{L}\left( 110\right) =3.82\cdot 10^{4}$; curve $L$ - $f_{L}\left(
110\right) =1.17\cdot 10^{4}$; curve $\protect\mu _{1}$ - $f_{L}\left(
110\right) =1.48\cdot 10^{5}$; curve $\protect\mu _{5}$ - $f_{L}\left(
110\right) =2.59\cdot 10^{4}$; curve $\protect\mu _{9}$ -$f_{L}\left(
110\right) =1.32\cdot 10^{4}$; curve $\protect\lambda _{0}$ - $f_{L}\left(
110\right) =6.95\cdot 10^{3}$; curve $\protect\lambda _{5}$ - $f_{L}\left(
110\right) =2.61\cdot 10^{4}$; curve $\protect\lambda _{9}$ - $f_{L}\left(
110\right) =5.7\cdot 10^{4}$. }
\label{fig4}
\end{figure}

We anticipated one feature, namely, the decrease in the amplitude of saw
teeth with increasing $x$ and even the disappearance of such teeth for the
Gaussian correlator. The sharp drops in conductivity in the points where the
number of the occupied minibands $S$ increases by 1 is explained by opening
of $S$ {\it new} scattering channels associated with interband transitions
in and from this newly opened miniband. Without the interband transitions,
the increase of $S$ by 1 results not in a sharp drop in $\sigma $, but in an
insignificant kink on the curve $\sigma \left( L\right) $ as it is shown in
the third of Refs.\cite{r2}. The interband transitions are described by the
off-diagonal components of the matrix of transition probabilities $%
W_{jj^{\prime }}$. With increasing $R/\Lambda _{F}$, these off-diagonal
(interband) transition probabilities go to zero though with different rate
for different types of the correlation function. The rate of decrease of the
interband transition probabilities as a function of $R/\Lambda _{F}$ for
different correlation functions is discussed in the Appendix. This rate is a
good predictor for observing the saw-like shape of $\sigma \left( L\right) $%
. The fastest decrease happens in the case of the Gaussian correlator; thus
the curve for the Gaussian correlator should be the smoothest and should
exhibit the smallest traces of the saw teeth. Therefore, the visibility of
the saw teeth on the experimental curve can be a clue to the form of the
correlation function. 
\begin{figure}[tbp]
\centerline{\epsfxsize=3.4in\epsfbox{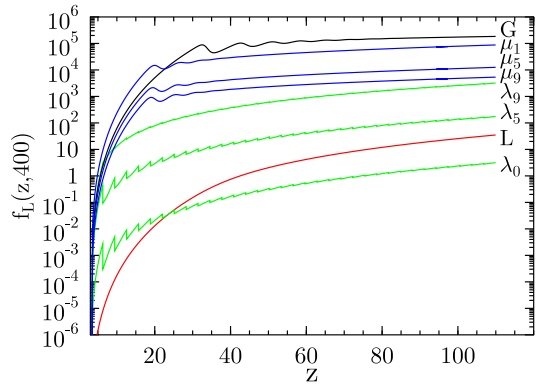}}
\caption{Functions $f_{L}\left( z,x=400\right) $ for the same 8 types of
correlators as in Figure 1. }
\label{fig5}
\end{figure}

What is completely unexpected is the appearance of a new type of oscillation
structure on $\sigma \left( L\right) $ in a limited range of $z$ for the
Gaussian and power-law correlators (curves $G$ and $\mu _{i}$ in Figures
5,6). It looks as if there is a transition between two distinct regimes with
several sharp oscillations in the transition range. The effect looks even
more striking in Figure 6 for the normalized curves which, in contrast to
Since Figure 5, is plotted in a linear scale. This new type of QSE requires
an explanation.

These new oscillations are not related to abrupt changes in the number of
occupied minibands $S\left( z\right) $: the oscillations are less sharp,
have a much larger period, and, most important, appear only in a limited
range of $z$ where the number of occupied minibands $S$\ is already large.
These new oscillations are observed for the correlators for which the
interband transitions are the smallest and the saw-like structure is
suppressed, namely for the Gaussian and power-law correlation functions. The
power spectrum for these correlators $\zeta \left( q\right) $\ goes to zero
exponentially\ at large $q$. Then one would expect that the off-diagonal
(interband) transition probabilities are exponentially small in comparison
with intraband scattering and that the conductivity can be well described by
the ''diagonal'' approximation $\left( \text{\ref{ee10}}\right) $ that does
not have an oscillation feature. This turns out not to be the case.

The oscillations are indeed related to off-diagonal (interband) scattering
probabilities $W_{jj^{\prime }}$. A qualitative explanation of the effect\
and an estimate of the peak positions are the following. Scattering by
surface inhomogeneities changes the tangential momentum by $\Delta q\sim 1/R$%
. According to the momentum conservation law, this scattering can cause the
interband transition $j\leftrightarrow j+1$ only when $q_{j}-q_{j+1}=\Delta
q\sim 1/R$. If the miniband index $j$ is relatively small and $q_{j}\sim
1/\lambda _{{\rm F}}$, then $q_{j}-q_{j+1}\sim \left(
q_{j}^{2}-q_{j+1}^{2}\right) \lambda _{{\rm F}}/2$. The energy conservation
requires that $q_{j}^{2}-q_{j+1}^{2}=\pi ^{2}\left( j+1\right)
^{2}/L^{2}-\pi ^{2}j^{2}/L^{2}\sim 2\pi ^{2}j/L^{2}$. The combination of
these conservation laws defines the peak positions $L_{j}$, which correspond
to the opening of robust interband transitions $j\leftrightarrow j+1$ and
which are given by equations $L_{j}^{2}\sim \pi ^{2}jR\lambda _{{\rm F}}$.
In dimensionless variables, this is equivalent to 
\begin{equation}
z_{j}\sim \pi \sqrt{jx}.  \label{eq35}
\end{equation}
Accordingly, with increasing film thickness $L$ the transition channel opens
first for the electrons in the lowest miniband $\epsilon _{1}\left( {\bf q}%
\right) $ with $j=1$. Note, that these are the grazing electrons which are
responsible for the dominant contribution to the conductivity. Thus, the
conductivity drops almost by half at the film thickness $z_{1}\sim \pi
x^{1/2}$ where $W_{12}$ becomes comparable to $W_{11}$ and the effective
cross-section doubles. At higher value of $L$, $z_{2}\sim \pi \left(
2x\right) ^{1/2}$, a new channel $W_{23}$ opens for the electrons from the
next miniband $j=2$ with $p_{x}=2\pi /L$ and the conductivity drops again,
and so on. The only difference is that the contribution of the electrons
from the higher minibands falls rapidly with an increase in the band index $j
$ and the drops in conductivity $\sigma \left( L\right) $, which are
associated with the opening of new scattering channels for electrons from
these minibands, become smaller and smaller. The number of the visible peaks
on the curve $\sigma \left( L\right) $ and their relative heights give a
good visual estimate of the number of ''important'' minibands and of their
relative contribution to the conductivity. With further increase in the film
thickness, when $L$\ becomes large, $L\gg R$, the change of momentum $\Delta
q\sim 1/R$ is sufficient to excite {\em all} interband transitions and the
ordinary QSE with the saw teeth at the points $z\sim \pi j$\ is restored.

The above explanation works for the films with the exponential decay of the
power spectrum of inhomogeneities in which the size of inhomogeneities $R$\
is well-defined. In the films with a non-exponential power spectrum of
inhomogeneities, {\em i.e.}, with a more uniform distribution of
inhomogeneities over the sizes $R$\ in momentum space, this\ new size effect
cannot be observed because the particles from all minibands can always find
the inhomogeneities of the right size that ensure the interband transitions
irrespective of what is the separation between the walls.

\begin{figure}[tbp]
\centerline{\epsfxsize=3.4in\epsfbox{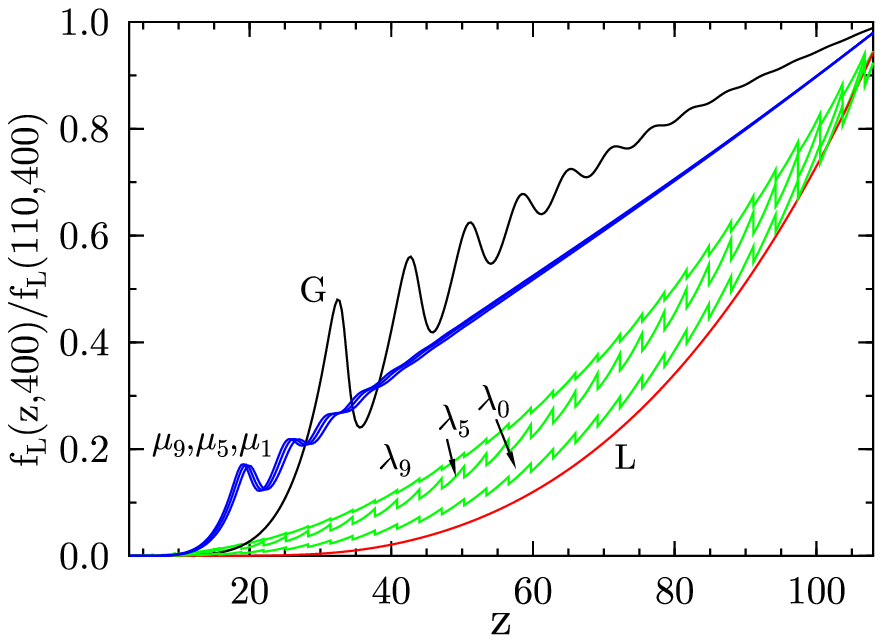}}
\caption{The same 8 functions $f_{L}\left( z,x=400\right) $ as in Figure 5
normalized by their value at $z=110$, $f_{L}\left( z\right) /f_{L}\left(
110\right) $. The normalization coefficients are: curve $G$ - $f_{L}\left(
110\right) =1.84\cdot 10^{5}$; curve $L$ - $f_{L}\left( 110\right) =35.0$;
curve $\protect\mu _{1}$ - $f_{L}\left( 110\right) =8.78\cdot 10^{4}$; curve 
$\protect\mu _{5}$ - $f_{L}\left( 110\right) =1.25\cdot 10^{4}$; curve $%
\protect\mu _{9}$ -$f_{L}\left( 110\right) =5.35\cdot 10^{3}$; curve $%
\protect\lambda _{0}$ - $f_{L}\left( 110\right) =3.16$; curve $\protect%
\lambda _{5}$ - $f_{L}\left( 110\right) =1.76\cdot 10^{2}$; curve $\protect%
\lambda _{9}$ - $f_{L}\left( 110\right) =3.21\cdot 10^{3}$. }
\label{fig6}
\end{figure}

More accurate explanation is the following. \bigskip The off-diagonal
elements $W_{jj^{\prime }}\;$are  functions of 
\[
\nu _{jj^{\prime }}=\left| q_{j}R/\hbar -q_{j^{\prime }}R/\hbar \right|
=x\left| \sqrt{1-\left( \pi j/z\right) ^{2}}-\sqrt{1-\left( \pi j^{\prime
}/z\right) ^{2}}\right| 
\]
and rapidly decrease with increasing $\nu _{jj^{\prime }},$ (see Appendix
A). In general, the off-diagonal $\nu _{jj^{\prime }}$ is large at large $R$
(or $x$) while the diagonal elements $\nu _{jj}=0$. However, for large $z$
(large $S$) some of the elements $\nu _{jj^{\prime }}$ with {\it small} $j,$
which are close to the main diagonal, could become small even for large $x$: 
\begin{equation}
\nu _{j,j+1}\left( j+1\ll z/\pi \right) \sim \frac{\pi ^{2}x}{2z^{2}}\left(
2j+1\right)   \label{d2}
\end{equation}
($j$ changes from $1$ to {\rm Int}$\left( z/\pi \right) $). Then at large $z$
the transitions $j\leftrightarrow j+1$ can become noticeable and Eqs. $%
\left( \text{\ref{ee6}}\right) $ become coupled. This coupling changes the
solution of transport equation and, therefore, conductivity. According to
Eqs. $\left( \text{\ref{ee6}}\right) $ the coupling between the minibands $j$
and $j+1$ becomes noticeable, $\tau _{j,j+1}^{-1}\sim \tau _{jj}^{-1}$, when 
\begin{equation}
W_{j,j+1}^{\left( 0\right) }\left( x,z\right) \sim W_{jj}^{\left( 0\right)
}\left( x,z\right) -W_{jj}^{\left( 1\right) }\left( x,z\right) .  \label{w1}
\end{equation}
At fixed $x$, Eq. $\left( \text{\ref{w1}}\right) $ can be considered as the
equation for the values of $z=z_{j}\left( x\right) $ at which one can
observe the opening of transitions $j\leftrightarrow j+1$. The opening of
such transition channels is accompanied by drops in conductivity. Since for
the Gaussian and power-law correlators the interband transition
probabilities $W_{jj^{\prime }}$ depend exponentially on parameters $\nu
_{jj^{\prime }}$, these drops in conductivity are sharp and deep as
illustrated in Figures 5,6. Solutions $z_{j}\left( x\right) $ of Eqs. $%
\left( \text{\ref{w1}}\right) $ are discussed in Appendix B. At $%
z=z_{1}\left( x\right) $, $W_{12}$ is the first of transition probabilities
to acquire the ''normal'' order of magnitude. At $z=z_{2}\left( x\right) $, $%
W_{23}$ becomes noticeable, then $W_{34}$, {\it etc.} The amplitudes of the
drops rapidly decrease with increasing $j$. In the end, when several
interband channels with $j\ll z/\pi $ are open, $\sigma \left( L\right) $
becomes smooth, but with a much lower slope than in its initial part. The
growth of transition probabilities for transitions $j\leftrightarrow j+2$
does not result in new oscillations in $\sigma \left( L\right) $. In the
points $z\left( x\right) $ where $W_{j,j+2}$ becomes large, $%
W_{j,j+2}^{\left( 0\right) }\sim W_{jj}^{\left( 0\right) }-W_{jj}^{\left(
1\right) }$, the states $j$ and $j+2$ are already strongly coupled via $%
W_{j,j+1}$\ and $W_{j+1,j+2}$.

According to Appendix B, Eq.(\ref{apb4}), the positions of the drops for
films with Gaussian surface inhomogeneities are similar to Eq. (\ref{eq35}):
\begin{equation}
z_{j}\left( x\right) \approx \frac{\pi }{2}\sqrt{\left( 2j+1\right) x}\left[
\ln \left( x\sqrt{2}\left( 1+1/j\right) \right) \right] ^{-1/4}.  \label{w3}
\end{equation}
The values $z_{j}\left( x=400\right) =33.4;43.6;51.8;58.9;....$ agree well
with the positions of the conductivity drops on curve 1 of Figures 5 and 6.

For the surface with the power-law correlations of inhomogeneities $\left( 
\text{\ref{ee2}}\right) $ the solution of Eq.$\left( \text{\ref{w1}}\right) $
with logarithmic accuracy [Appendix B, Eq.(\ref{apb9})] again resembles Eq. (%
\ref{eq35}):
\begin{eqnarray}
z_{j}\left( x\right)  &=&\pi \sqrt{\left( 2j+1\right) x/4\nu },  \label{w4}
\\
\nu  &\sim &\ln \left[ x\left( 1+1/j\right) \left\{ 2\ln \left[ x\left(
1+1/j\right) \right] \right\} ^{\mu /2+1/4}\right] .  \nonumber
\end{eqnarray}
This expression is barely sensitive to $\mu $. This almost complete
independence of the peak positions from $\mu $ can be clearly seen in Figure
6.

The difference between this new type of size effect and the usual saw-like
QSE is dramatic. The saw-like drops in conductivity for usual QSE occur in
the points $z=k\pi $ with integer $k$ and are direct consequence of
quantization of momentum in thin films. The interband transitions are not
germane to the existence or positions of this QSE and are responsible only
for the amplitude of the conductivity oscillations. The drops in
conductivity are equidistant with the period $\pi $ along the $z$ axis, {\it %
i.e, }are equidistant as a function of the film thickness. In contrast to
this, the new QSE oscillations in Figures 5,6 are not related directly to
the quantization of momentum and are a consequence of the exponential
opening of interband transitions between minibands with small quantum
numbers at certain values of the film thickness. The transitions in and out
of higher minibands remain suppressed. (In some sense, the effect resembles
magnetic breakthrough between separated parts of the Fermi surface in high
magnetic fields). The peaks are roughly equidistant if plotted against $%
z^{2} $; weak deviation from periodicity is due to a logarithmic terms in
Eqs. $\left( \text{\ref{w3}}\right) ,\left( \text{\ref{w4}}\right) $. The
period of the new QSE is much larger than for the usual QSE. The large
period of oscillations can open the way to direct observation of QSE in
transport measurements in metal films in which usual QSE has atomic period
and can hardly be observed. There is a strong possibility that the
conductivity oscillations reported in the last of Refs. \cite{qse1} are
actually this new type of QSE.

The initial part of the curves $G$, $\mu _{i}$ in Figures 5,6 for $\sigma
\left( L\right) $ is described analytically by Eq.$\left( \text{\ref{ee10}}%
\right) $ with appropriate values of $W$ from Appendix A. This curve is
close to the power law $\sigma \propto L^{\left( 5+\alpha \right) }$ (small $%
\alpha $ depends on $x$) and to experimental data of the third Ref. \cite
{qse1}. After the region of new QSE oscillations, the curves are again
smooth, but with a much smaller tangent. We do not have an analytical
description for this regime. The numerical approximation can be done equally
well by either $\sigma =A+B\cdot L^{1+\beta }$ with small $\beta $ ($\beta $
also depends on $x$) or a quadratic expression $a+b\cdot L+c\cdot L^{2}$.
This behavior explains the experimental data \cite{or1} and the last Ref. 
\cite{conduct1}. As a result, the power-law dependence of $\sigma \left(
L\right) $ is qualitatively different for ultrathin and more thicker films.
This type of behavior is different from the earlier studied behavior of $%
\sigma \left( L\right) $ at small $x=q_{F}R\ll 1$. \cite{fish2,arm2,qq17}. 
\begin{figure}[tbp]
\centerline{\epsfxsize=3.4in\epsfbox{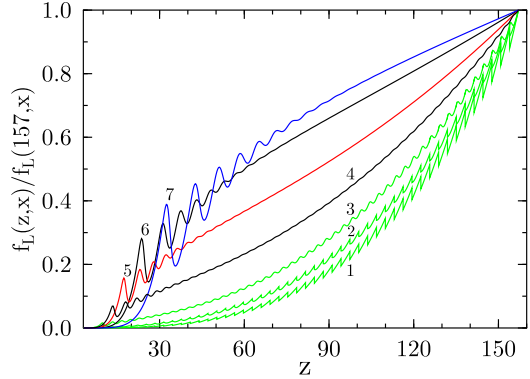}}
\caption{Functions $f_{L}\left( z,x=const\right) $ for Gaussian correlation
of surface inhomogeneities normalized by their value at $z=157$, $%
f_{L}\left( z\right) /f_{L}\left( 157\right) $. The values of $x$ and
normalization coefficients are: curve 1 - $x=1$, $f\left( 157\right)
=6.9\cdot 10^{6}$; curve 2 - $x=10$, $f\left( 157\right) =9.9\cdot 10^{4}$%
;curve 3 - $x=25$, $f\left( 157\right) =4.6\cdot 10^{4}$;curve 4 - $x=55$, $%
f\left( 157\right) =3.8\cdot 10^{4}$; curve 5 - $x=100$, $f\left( 157\right)
=4.75\cdot 10^{4}$; curve 6 - $x=200$, $f\left( 157\right) =9.1\cdot 10^{4}$%
; curve 7 - $x=400$, $f\left( 157\right) =2.3\cdot 10^{5}$. }
\label{fig7}
\end{figure}

The initiation of this new type of oscillations with a large period can be
seen on the initial part of curve $G$ in Figure 2 for $x=10.$ With growing $%
z $ these new oscillations get overtaken by the standard QSE. The transition
from standard to new QSE is illustrated in Figure 7 that contains normalized
''curves $G$'' for the Gaussian inhomogeneities, $f_{L}\left(
z,x=const\right) /f_{L}\left( z=157\right) $, for $x=1;10;25;55;100;200;400$%
. It is clear from these curves how usual QSE is replaced by new
oscillations with increasing $x$. The ''transitional'' curve for $x=55$ is
especially interesting: it shows new QSE at smaller $z$ and a restoration of
standard QSE at higher $z$. This restoration occurs when a noticeable number
of interband transitions become open at higher $z$. It seems that such
restoration does not happen on curves $x>50$. This impression is wrong. Such
restoration indeed occurs for curves $x=100;200;400$, but at values of $z$
that are much larger than those in the Figure. At very large $x$, all curves 
$f_{L}\left( z,x=const\right) $ consist of four parts: rapid increase at
small $z$, region of new QSE oscillations, smooth monotonic part, and the
region of relatively smooth standard QSE oscillations at the largest values
of $z$. With increasing $x$, the amplitude of new QSE oscillations and the
length of region separating new and old QSE increase rapidly.

\subsection{Dependence on the correlation radius}

The dependence of the conductivity on the correlation radius of surface
inhomogeneities, $\sigma \left( R\right) $, is best illustrated by the
function $f_{R}\left( y,z=const\right) $, Eq. $\left( \text{\ref{c4}}\right) 
$. Since the number of the occupied minibands $S$ does not depend on the
correlation radius of inhomogeneities, the curves $f_{R}\left( y\right) $ at
constant $z$ do not exhibit the saw-like structure. Instead, the two main
features are the presence of the minimum in $f_{R}\left( y\right) $ and the
step-like structure that corresponds to the oscillations in Figures 5,6. 
\begin{figure}[tbp]
\centerline{\epsfxsize=3.4in\epsfbox{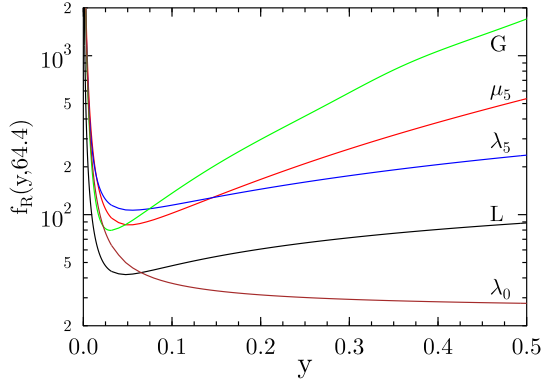}}
\caption{Function $f_{R}\left( y,z=64.4\right) $, Eq.$\left( \text{\ref{c4}}%
\right) $, near the minimum at $yz\sim 1$ for various surface correlators.
The labeling of the curves is explained in the end of Sec. IVA. }
\label{fig8}
\end{figure}

The scattering of fermions by surface inhomogeneities is most effective at $%
R/\Lambda _{F}\sim 1$, {\it i.e., }at $y\sim 1/z$. This leads to a minimum
of the conductivity $\sigma \left( R\right) $ at such values of $y$. At $%
R/\Lambda _{F}\ll 1$ the particle wavelength is much larger than the size of
surface inhomogeneities and the scattering is almost specular and does not
contribute to the formation of the mean free path. In the opposite limit $%
R/\Lambda _{F}\gg 1$ the walls are flat on the particle length scale and
surface scattering also does not limit the effective mean free path.
Therefore, at $z=const$ the conductivity $\sigma \left( R\right) $ for
non-divergent correlators is infinite in both limits $y\rightarrow 0$ and $%
y\rightarrow \infty $ with a minimum around $y\sim 1/z$. The curves $%
f_{R}\left( y\right) $ close to this minimum are plotted for different
correlators in Figure 8 ($z=64.4$; the labeling of the curves is explained
in the end of Sec. IVA). It is important that the position of the minimum,
its width, and even the order of magnitude of the function $f_{R}\left(
y\right) $ in the minimum are roughly the same for all types of surface
correlators. This is, probably, the most universal feature of the system.
The only correlator that does not display a well-defined minimum is $\left( 
\text{\ref{ee3}}\right) $\ with $\lambda =0$ (the Lorentzian in momentum
space; curve $\lambda _{0}$). This feature is related to the logarithmic
divergence of this correlator in ''real'' space. This feature is especially
interesting because the surfaces with such inhomogeneities were observed in
experiment \cite{fer1}.

The drops in $\sigma \left( L\right) $\ at large $z=z_{j}\left( x\right) $,
which are analyzed in the previous Section (Figures 5,6), correspond to the
points $y_{j}\left( z\right) $ on the curves $f_{R}\left( y\right) $. The
positions of these points $y_{j}\left( z\right) $ are implicitly determined
by Eq. $\left( \text{\ref{w3}}\right) $ and $\left( \text{\ref{w4}}\right) $
for the Gaussian and power-law correlations provided that $x=yz$. These
values of $y$ are far away to the right from the minimum in the curves $%
\sigma \left( R\right) $ and cannot be presented in the same figures. The
feature that corresponds to the oscillations from the previous Section is
clearly seen as a set of steps in Figure 9 for the same value of $z$ as in
Figure 8, $z=64.4$ on curves $G$ and $\mu _{5}$ for Gaussian and power-law
inhomogeneities. For the surfaces with the Gaussian inhomogeneities , the
first interband transition $W_{12}$ becomes visible for $z=64.4$ at $%
y_{1}\sim 25$, the next one at $y_{2}\sim 14$, and so on. At these values of 
$y$ one can see well-pronounced steps on the curve $G$ in Figure 9. The same
feature, though barely discernible, is also observed for the power-law
correlator $\mu _{5}$.

For comparison, curves $L,$ $\lambda _{0},$ and $\lambda _{5}$ do not
exhibit any anomalies. Interestingly, the curve for the Lorentzian
inhomogeneities is the only one that decreases with increasing $y$ after the
initial increase at small $y$ (Figure 8). How is this feature related to the
peculiarities of the Lorentzian that have been discussed in Section II is
unclear. The curve $\lambda _{0}$ remains essentially flat. 
\begin{figure}[tbp]
\centerline{\epsfxsize=3.4in\epsfbox{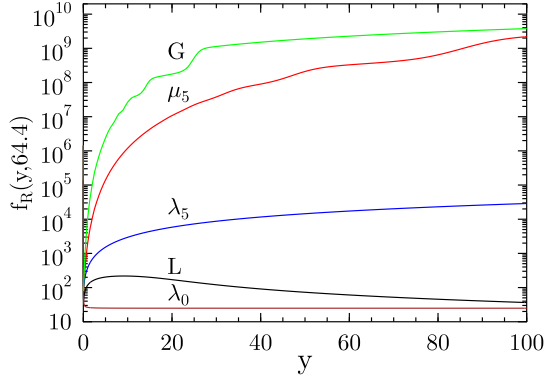}}
\caption{The same functions $f_{R}\left( y,z=64\right) $, Eq.$\left( \text{%
\ref{c4}}\right) $, as in Figure 8 at larger values of $z$. The labeling of
the curves is explained in the end of Sec. IVA. }
\label{fig9}
\end{figure}

\subsection{Dependence on the Fermi momentum and density of fermions}

The dependence of the conductivity $\sigma $ on the density of fermions $N$
or their Fermi momentum $q_{F}$ is best displayed by the function $%
f_{N}\left( z\right) $ at constant $y=R/L$, see Eq. $\left( \text{\ref{c5}}%
\right) $. This dependence $\sigma \left( N\right) $ is similar to $\sigma
\left( L\right) $. The function $\sigma \left( N\right) $ exhibits a clear
saw-like structure of usual QSE at not very high $y$ for {\it all}
correlators. With increasing $y$, the saw teeth disappear first for the
Gaussian correlator $G$, then for the power-law correlators $\mu _{i}$, but
persist for the power-law correlators in momentum space $\lambda _{i}$.
Instead, at large $y$ the functions $f_{N}\left( z,y=const\right) $ for
Gaussian and power-law inhomogeneities exhibit new type of QSE oscillations
similar to that for $f_{L}\left( z,x=const\right) $ in Sec. IVB. The
positions of these oscillations can be found from Eqs.$\left( \text{\ref{w3}}%
\right) ,\left( \text{\ref{w4}}\right) $ after substitution $x=yz$. 
\begin{figure}[tbp]
\centerline{\epsfxsize=3.4in\epsfbox{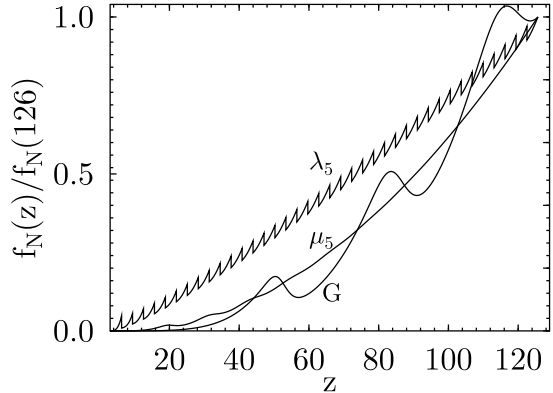}}
\caption{Normalized function $f_{N}\left( z,y=20\right) $ Eq.$\left( \text{%
\ref{c5}}\right) $, $f_{N}\left( z\right) /$ $f_{N}\left( z=126\right) $,
for three surface correlators. The normalization coefficients are: curve $G$
- $f_{L}\left( 126\right) =1.1\cdot 10^{9}$; curve $\protect\mu _{5}$ - $%
f_{L}\left( 126\right) =4.5\cdot 10^{7}$; curve $\protect\lambda _{5}$ - $%
f_{L}\left( 126\right) =1.4\cdot 10^{4}$. }
\label{fig10}
\end{figure}

This effect is illustrated in Figure 10 (the labeling of the curves is
explained in the end of Sec. IVA). The figure presents functions $%
f_{N}\left( z,y=20\right) $, Eq.$\left( \text{\ref{c5}}\right) $, for the
Gaussian (curve $G$) and power-law ($\mu =0.5$; curve $\mu _{5}$)
correlators, and for the correlator with a power-law power spectrum ($%
\lambda =0.5$; curve $\lambda _{5}$). To compensate for different orders of
magnitude of the data for these correlators, the functions are normalized by
their values at $z=126$, $f_{N}\left( z\right) /f_{N}\left( z=126\right) $.
Curve $\lambda _{5}$ exhibits a saw-like behavior typical to the usual QSE
with period $\pi $ along the $z$-axis. Curves $G$ and $\mu _{5}$ exhibit new
QSE oscillations with a much larger period.

\subsection{Interwall correlation of inhomogeneities and quantum size effect}

Surprisingly, the possibility of interwall correlation of surface
inhomogeneities gives an interesting insight into usual and new QSE and
provides an additional proof for our explanation of QSE oscillations
reported above. The study of the effect of interwall correlation of
inhomogeneities has been initiated in Ref.\cite{arm1} for Gaussian
correlations. Below we supplement those results for other types of surface
correlators with an emphasis on new QSE.

To decrease the number of parameters, we assume that, as in Ref.\cite{arm1},
the correlation functions of inhomogeneities on both walls $\zeta _{11}$ and 
$\zeta _{22}$ are given by the same function, $\zeta _{11}\left( s\right)
=\zeta _{22}\left( s\right) =\zeta \left( s\right) $. The structure of the
interwall correlator of inhomogeneities $\zeta _{12}\left( s\right) $ is
assumed to be the same as for the intrawall correlations with the same
correlation radius $R$. However, the amplitude $a$ of the interwall
correlations is different from the intrawall ones, 
\begin{equation}
\zeta _{11}=\zeta _{22}=\zeta \left( s\right) ,\ \zeta _{12}\left( s\right)
=a\zeta \left( s\right) .  \label{ii1}
\end{equation}
To compare the effect of such interwall correlations for different classes
of the function $\zeta \left( s\right) $, we calculate the relative change
of conductivity $\sigma $ ({\it i.e., }functions $f_{L},f_{R},f_{N}$) caused
by introduction of such correlations 
\begin{equation}
\phi ^{\left( a\right) }=\frac{f^{\left( a\right) }-f}{f},  \label{i1}
\end{equation}
where $f^{\left( a\right) }$ and $f$ are the functions $f_{L,R,N}$
calculated with and without interwall correlations. An additional benefit is
that the functions $\phi ^{\left( a\right) }$ for all types of correlators
are automatically normalized thus eliminating a difference by orders of
magnitude between the functions $f_{L,R,N}$ for different types of
correlation functions.
\begin{figure}[tbp]
\centerline{\epsfxsize=3.4in\epsfbox{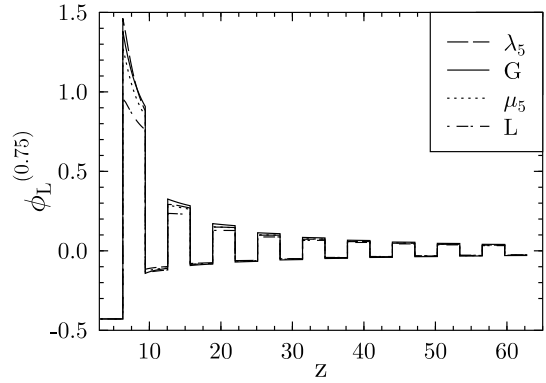}}
\caption{Relative change $\protect\phi _{L}^{\left( 0.75\right) }$, Eq.$%
\left( \text{\ref{i1}}\right) ,$ of the function $f_{L}\left( z,x=1\right) ,$
Eq. $\left( \text{\ref{c3}}\right) ,$\ for the interwall correlation
amplitude $\left( \text{\ref{ii1}}\right) $ $a=0.75$ for various correlation
functions of surface inhomogeneities. The labeling of the curves is
explained in the end of Sec. IVA. All the curves exhibit almost identical
oscillations as it should be for well-developed usual QSE. }
\label{fig11}
\end{figure}

In the presence of such interwall correlations, the transition probabilities 
$W_{jj^{\prime }}\left( {\bf q},{\bf q}^{\prime }\right) $ $\left( \text{\ref
{rrr3}}\right) $\ become proportional, in accordance with \cite{arm1}, to 
\begin{equation}
2\left[ 1+a\left( -1\right) ^{j+j^{\prime }}\right] \zeta \left( \left| {\bf %
q}_{j}-{\bf q}_{j^{\prime }}^{\prime }\right| \right) .  \label{i2}
\end{equation}
The most interesting effects of the interwall correlations are related to
the oscillating structure of the term with $a$ in Eq. $\left( \text{\ref{i2}}%
\right) $. If the interband transition probabilities $W_{j\neq j^{\prime
}}\left( {\bf q},{\bf q}^{\prime }\right) $ are large, {\it i.e}, if $\zeta
\left( \left| {\bf q}_{j}-{\bf q}_{j^{\prime }}^{\prime }\right| \right) $
is not small for $j^{\prime }\neq j$, then the contribution of the term with 
$a$\ in $\left( \text{\ref{i2}}\right) $ has a different sign for different $%
W_{jj^{\prime }}$ depending on whether $j+j^{\prime }$\ is even or odd. This
should result in an oscillating structure of the function $\phi ^{\left(
a\right) }$ $\left( \text{\ref{i1}}\right) $ as a function of the number of
occupied minibands $S$, {\it i.e.,} as a function of film thickness $L$ (the
existence of such oscillations was first reported in Ref.\cite{arm1} for
Gaussian inhomogeneities). The period of such oscillations should be equal
to that for standard QSE and their amplitude should decrease rapidly with
increasing $L$. Since our explanation of the standard QSE ties it to large
interband transitions, the oscillation nature of the function $\phi ^{\left(
a\right) }$ $\left( \text{\ref{i1}}\right) $ should exist in the same range
of parameters as the standard QSE. In accordance with Sec. IVB, these
oscillations should be noticeable for the function $\phi _{L}^{\left(
a\right) }\left( z,x=const\right) $ at small $x$\ for {\it all} types of
surface correlators. This is illustrated in Figure 11 ($x=1$) for the
correlators $G$, $L$, $\lambda _{5}$, $\mu _{5}$. The figure is plotted for $%
a=0.75$. The similarity of the functions $\phi _{L}^{\left( 0.75\right)
}\left( z,x=1\right) $ is striking, but not surprising. The flat part of all
curves at small $z$ is explained below. At higher values of $x$, the
interband transitions (off-diagonal $W_{jj^{\prime }}$ $\left( \text{\ref{i2}%
}\right) $) become more and more suppressed. When the interband transitions
become negligible, the only non-zero scattering probabilities are diagonal $%
W_{jj}$ that are proportional to $2\left[ 1+a\right] \zeta \left( \left| 
{\bf q}_{j}-{\bf q}_{j}^{\prime }\right| \right) $ Eq.$\left( \text{\ref{i2}}%
\right) $. Since {\em all} $W_{jj}$\ are scaled by the same factor $1+a$\
and the conductivity is inversely proportional to $W$, the function $\phi
_{L}^{\left( a\right) }\left( z\right) $ in the absence of the interband
transition becomes a constant, 
\begin{equation}
\phi _{L}^{\left( a\right) }\left( z\right) =\frac{1}{1+a}-1.  \label{i5}
\end{equation}
If $a=0.75$, the value of this constant is $\phi _{L}^{\left( 0.75\right)
}\left( z\right) =-3/7$. Eq.$\left( \text{\ref{i5}}\right) $ also describes
the initial part of all curves $\phi _{L}^{\left( a\right) }\left( z\right) $%
\ for all values of $x$\ at small $z$\ when only the first miniband is
occupied, $S=j=j^{\prime }=1$. This explains all curves in Figure 11 have
identical flat pat at small $z$. 
\begin{figure}[tbp]
\centerline{\epsfxsize=3.4in\epsfbox{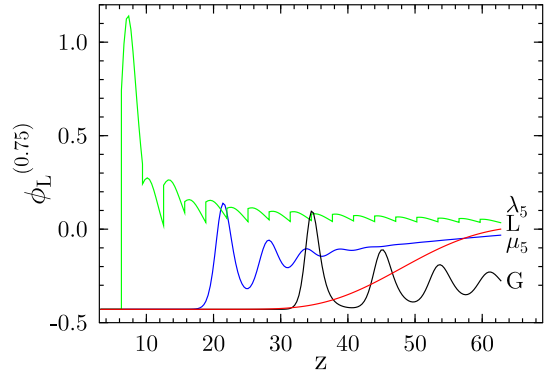}}
\caption{Relative change $\protect\phi _{L}^{\left( 0.75\right) }$, Eq.$%
\left( \text{\ref{i1}}\right) ,$ of the function $f_{L}\left( z,x=400\right) 
$ $\left( \text{\ref{c3}}\right) $\ for the interwall correlation amplitude $%
\left( \text{\ref{ii1}}\right) $ $a=0.75$ for four correlation functions of
surface inhomogeneities. The labeling of the curves is explained in the end
of Sec. IVA. Curve $\protect\lambda _{5}$ exhibits oscillations in
accordance with usual QSE for curve $\protect\lambda _{5}$ in Figure 6.
Curves $G$, $L$, and $\protect\mu _{5}$ are flat at small $z,$ $\protect\phi %
_{L}^{\left( 0.75\right) }=-3/7$, Eq.$\left( \text{\ref{i5}}\right) $.
Oscillations on curves $G$ and $\protect\mu _{5}$ confirm the explanation of
the new QSE as an exponential appearance of transitions $%
j\longleftrightarrow j+1$ at certain values of $z$. }
\label{fig12}
\end{figure}
Figure 12 illustrates $\phi _{L}^{\left( a\right) }\left( z,x=const\right) $
at $x=400$ and $a=0.75$\ for several correlators. At this value of $x$, the
exponential correlator $\lambda _{5}$ $\left( \text{\ref{e2}}\right) $
exhibits, according to the results and explanation of Sec. IVB, the usual
QSE. Therefore, the function $\phi _{L}^{\left( 0.75\right) }\left(
z,x=400\right) $\ for this correlator should have an oscillation structure;
this is clearly seen in Figure 12. The Gaussian and power-law correlators $G$
and $\mu _{5}$, according to Sec. IVB, ensure the absence of interband
transitions at small and moderate $z$ where the function $\phi _{L}^{\left(
0.75\right) }=-3/7$ in Figure 12. Our explanation for the new type of QSE in
Sec. IVB is an abrupt sequential appearance of noticeable interband
transitions $W_{12}$, $W_{23}$, $W_{34}$, {\it etc.} at certain values of $%
z=z_{j}$. Since for the term with $a$ in Eq.$\left( \text{\ref{i2}}\right) $
is negative for all transitions $j^{\prime }=j\pm 1$, one should observe
spikes in conductivity and, therefore, in the function $\phi _{L}^{\left(
a\right) }$, at $z=z_{j}$. In some sense, Figure 12 provides the best
illustration for our explanation of new QSE.

Figure 12 also provides an insight into anomalous behavior of conductivity
for Lorentzian correlation of inhomogeneities $\left( \text{\ref{eee2}}%
\right) $, curve $L$. At $z<30$, the interband transitions are suppressed
and $\phi _{L}^{\left( 0.75\right) }=-3/7$. At higher $z$, the interband
transitions become more noticeable and start increasing, but very slowly.
Why does the curve remain smooth when a sufficient number of transitions is
already visible, is still a puzzle. A possible explanation is that
oscillations should appear only at very large $S$ (or $z$) when their
amplitude should be vanishingly small.

\section{Summary and conclusions}

In summary, we compared the behavior of conductivity for various types of
surface correlators in a wide range of parameters. The following conclusions
can be important when analyzing the experimental data or discussing
theoretical predictions.

\begin{itemize}
\item  The rough shapes of the curves of the transport coefficients are
similar at small and moderate $R$ for {\em all} types of correlators though
the orders of magnitude of the transport coefficients and more fine details
of the curves can be different. To make any definite conclusions from the
rough shapes of the experimental curves, one should have at least some idea
of the type of the correlation function of surface inhomogeneities and/or
the value of the correlation radius $R$ and the amplitude of inhomogeneities 
$\ell $. Since $\ell $ plays the role of a scaling parameter, getting the
values of parameters of surface inhomogeneities from experimental data on
transport without any additional information on the correlation of
inhomogeneities could result in mistakes by orders of magnitude. In the same
way, the use of the wrong correlator in theoretical calculations could
result in absolutely wrong predictions without evoking any warning signals
from comparison of the rough shapes of experimental and theoretical curves.

\item  The most universal feature is the shape of the curves and order of
magnitude of $\sigma \left( R\right) $ near the minimum at $R/\Lambda
_{F}\sim 1$. This minimum allows experimental evaluation of the correlation
length of surface inhomogeneities $R$ without any assumptions about the type
of the correlation function.

\item  The shape of the curves $\sigma \left( L\right) $, $\sigma \left(
N\right) $, and $\sigma \left( R\right) $ becomes very sensitive to the type
of surface correlator at large correlation radius of inhomogeneities $R$.
Experimentally, this is important for better quality films (see, for
example, in Ref. \cite{coupl1}) in which STM and other usual methods are not
well-suited for the study of\ the long-range behavior of the thickness
fluctuations. Here transport measurements can be used as a good alternative
for identification and analysis of the thickness fluctuations.

\item  The underlying reason is very high sensitivity of coupling between
quantum well states with {\em low} quantum numbers to film thickness and the
long-range behavior of the thickness fluctuations. This phenomenon is quite
general and should lead to observable effects not only in metal films, but
for other types of quantum wells such as semiconductor films or quantum wave
guides \cite{mar1}.

\item  The persistence of the saw-like dependence of the transport
coefficients on the thickness of the film, Fermi momentum, or the density of
fermions should signal the long range nature of the surface correlations in
momentum space $\zeta \left( q\right) $. The observation of the saw-like
structure for $R>L$ is a distinct signature of the power-law decay of the
power spectral density function $\zeta \left( q\right) $ though, by itself,
is insufficient to make conclusions about the index in this power law. The
easy suppression of the saw-like behavior points at the exponential decay of
the power spectral density. The rate of this suppression is significantly
different for simple exponential and Gaussian decays of $\zeta \left(
q\right) $.

\item  Thickness fluctuations with Gaussian correlations and correlations
with exponential power spectrum lead to a new type of QSE in $\sigma \left(
L\right) $, $\sigma \left( N\right) $, and $\sigma \left( R\right) $ for
surface inhomogeneities of a relatively large size $R$. This new QSE
produces large oscillations in $\sigma \left( L\right) $ and $\sigma \left(
N\right) $ and steps in dependence $\sigma \left( R\right) $. The spacing
between these new QSE anomalies provides important direct information on the
correlation parameters of inhomogeneities. The peaks are almost equidistant
if plotted against $z^{2}$.

\item  In contrast to the usual saw-like QSE, the new QSE oscillations are
not related directly to the quantization of momentum and are a consequence
of the exponential opening of interband transitions between minibands with
small quantum numbers at certain values of the film thickness. In some
sense, the effect is reminiscent of magnetic breakthrough that describes the
opening of transitions between disconnected parts of the Fermi surface.

\item  Large period of new QSE oscillations opens the way to direct
observation of QSE in conductivity of quantized metal films and may be
responsible for experimental data in the second Ref. \cite{conduct1}. An
additional experimental signature should be the appearance of these new QSE
oscillations only at relatively large values of the thickness of quantized
metal films.

\item  The Gaussian correlation of inhomogeneities affects particle
transport in a unique way. First, the values of the transport coefficient
are, except for the smallest correlation radii, larger than for other,
slower correlators by orders of magnitude. This is explained by this
correlator having the shortest tails resulting in the least effective
scattering. Second, this type of correlation does not exhibit a saw-like
dependence of the transport coefficients on the system parameters except for
small correlation radii $R$. Third, this type of correlation of the surface
inhomogeneities leads to the above-mentioned new type of large-scale
oscillations of the transport coefficients. The combination of these
features can make the Gaussian correlator readily identifiable in transport
experiments.

\item  The Lorentzian correlation of inhomogeneities in configuration space
is also readily identifiable by several abnormal features. The combination
of these features could be another manifestation of an ''unphysical'' nature
of this correlator. If possible, this correlator should be avoided in
theoretical and computational models. A power-law correlator $\left( \text{%
\ref{ee2}}\right) $ with small index $\mu $ can serve as a good replacement
in the calculations.

\item  The results explain the observed difference in power-law regimes of
the thickness dependence of the conductivity $\sigma \left( L\right) $
between ultrathin and more thicker films.

\item  The relative contribution of the interwall correlation of surface
inhomogeneities strongly depends on the type of QSE. For usual QSE, the
contribution of the interwall correlations is a rapidly decaying oscillation
function of the film thickness. For QSE of the new type, this contribution
is constant in a wide range of small and moderate thicknesses, and becomes
an oscillating function with a big period in a limited range of large
thicknesses.
\end{itemize}

\section{Acknowledgments}

The work is supported by NSF grant DMR-0077266.

\section{Appendix A. Transition probabilities}

Various correlation functions from Section II allow different degrees of
analytical calculations of the scattering probabilities. The angular
harmonics of the correlation function $\zeta \left( \left| {\bf q-q}^{\prime
}\right| \right) $ in the transport equation $\left( \text{\ref{ee6}}\right) 
$ are defined as 
\begin{eqnarray}
\zeta \left( \left| {\bf q-q}^{\prime }\right| \right) &=&\frac{1}{2}\zeta
^{\left( 0\right) }\left( q,q^{\prime }\right) +\sum_{s=1}^{\infty }\zeta
^{\left( s\right) }\left( q,q^{\prime }\right) \cos \left( s\chi \right) ,
\label{ap0} \\
\zeta ^{\left( s\right) } &=&\frac{1}{\pi }\int_{0}^{2\pi }\zeta (\chi )\cos
\left( s\chi \right) \,d\chi  \nonumber
\end{eqnarray}
where $\chi $ is the angle between the $2D$ vectors ${\bf q}$\ and ${\bf q}%
^{\prime }$.

The harmonics for the Gaussian correlator $\left( \text{\ref{a2}}\right) $
are 
\begin{eqnarray}
\zeta ^{\left( 0\right) }\left( q_{j},q_{j^{\prime }}\right) &=&4\pi \ell
^{2}R^{2}\left[ e^{-QQ^{\prime }}I_{0}\left( QQ^{\prime }\right) \right]
e^{-\left( Q-Q^{\prime }\right) ^{2}/2},  \label{ap1} \\
\zeta ^{\left( 1\right) }\left( q_{j},q_{j^{\prime }}\right) &=&4\pi \ell
^{2}R^{2}\left[ e^{-QQ^{\prime }}I_{1}\left( QQ^{\prime }\right) \right]
e^{-\left( Q-Q^{\prime }\right) ^{2}/2}  \nonumber
\end{eqnarray}
where $Q=q_{j}R$, $Q^{\prime }=q_{j^{\prime }}R$. Note, that in Refs.\cite
{arm1,arm2,arm3} we used equivalent expressions with hypergeometric
functions instead of modified Bessel functions. Expressions in square
brackets in Eqs. $\left( \text{\ref{ap0}}\right) $\ are smooth functions of $%
Q$ and $Q^{\prime }$. The exponential coefficients, $\exp \left[ -\left(
Q-Q^{\prime }\right) ^{2}/2\right] $, on the other hand, are rapidly going
to zero for large $qR$ if $q_{j}\neq q_{j^{\prime }}$. This explains why the
off-diagonal scattering probabilities $W_{jj^{\prime }}$\ are much smaller
than the diagonal ones at large $qR$. Such a drastic difference between
interband and intraband scattering probabilities is a unique feature of the
Gaussian correlator. The physical consequences are discussed in Section IV.

For the exponential correlator $\left( \text{\ref{e2}}\right) $\ the
harmonics are

\begin{eqnarray}
\zeta ^{\left( 0\right) }\left( q_{j},q_{j^{\prime }}\right) &=&\frac{8\ell
^{2}R^{2}E\left( \Omega \right) }{\left[ 1+\left( Q-Q^{\prime }\right) ^{2}%
\right] \sqrt{1+\left( Q+Q^{\prime }\right) ^{2}}}  \label{ap2} \\
\zeta ^{\left( 1\right) }\left( q_{j},q_{j^{\prime }}\right) &=&\frac{4\ell
^{2}R^{2}}{QQ^{\prime }}\frac{\left( 1+Q^{2}+Q^{\prime 2}\right) E\left(
\Omega \right) -\left( 1+\left( Q-Q^{\prime }\right) ^{2}\right) K\left(
\Omega \right) }{\left[ 1+\left( Q-Q^{\prime }\right) ^{2}\right] \sqrt{%
1+\left( Q+Q^{\prime }\right) ^{2}}},  \nonumber \\
\Omega &=&2\sqrt{QQ^{\prime }/\left[ 1+\left( Q+Q^{\prime }\right) ^{2}%
\right] },  \nonumber
\end{eqnarray}
where $E$ and $K$ are complete elliptic integrals. Here the diagonal and
off-diagonal transition probabilities (probabilities of the intraband and
interband scattering) differ mainly by the terms $1+\left( Q-Q^{\prime
}\right) ^{2}$ in denominator that are insignificant in comparison with the
exponential factors for the Gaussian correlator above. The physical
consequences are discussed in Section IV.

The power-law $\left( \text{\ref{ee2}}\right) $ correlation functions
correspond to

\begin{eqnarray}
\zeta ^{(0)} &=&4\ell ^{2}R^{2\,}\sum_{m=0}^{\infty }\left( \mu +m\right) 
\frac{K_{\mu +m}\left( Q_{\max }\right) }{Q_{\max }^{\mu }}\frac{I_{\mu
+m}\left( Q_{\min }\right) }{Q_{\min }^{\mu }}  \label{ap3} \\
&&\times \int_{0}^{2\pi }C_{m}^{\mu }(\cos \phi )\left[ Q^{2}+Q^{\prime
2}-2QQ^{\prime }\cos \phi \right] ^{\mu }\,d\phi  \nonumber \\
\zeta ^{(1)} &=&4\ell ^{2}R^{2\,}\sum_{m=0}^{\infty }\left( \mu +m\right) 
\frac{K_{\mu +m}\left( Q_{\max }\right) }{Q_{\max }^{\mu }}\frac{I_{\mu
+m}\left( Q_{\min }\right) }{Q_{\min }^{\mu }}  \nonumber \\
&&\times \int_{0}^{2\pi }C_{m}^{\mu }(\cos \phi )\left[ Q^{2}+Q^{\prime
2}-2QQ^{\prime }\cos \phi \right] ^{\mu }\cos \phi \,d\phi  \nonumber
\end{eqnarray}
where $C_{m}^{\mu }$ are the ultraspherical (Gegenbauer) polynomials, and $%
Q_{\max }=\max (Q,Q^{\prime })$ and $Q_{\min }=\min (Q,Q^{\prime })$. The
off-diagonal transition probabilities disappear exponentially at large $%
\left| Q-Q^{\prime }\right| $, approximately as $\left( \left| Q-Q^{\prime
}\right| \right) ^{\mu -1/2}\exp \left( -\left| Q-Q^{\prime }\right| \right) 
$, {\it i.e., }much slower than for the Gaussian correlator $\left( \text{%
\ref{ap1}}\right) $ but faster than for the correlator $\left( \text{\ref
{ap2}}\right) $.

The integrals in Eqs.$\left( \text{\ref{ee2}}\right) $ can be simplified for
the Lorentzian correlator$:$

\begin{eqnarray}
\zeta ^{(0)}(Q,Q^{\prime }) &=&8\pi \ell ^{2}RK_{0}\left( Q_{\max }\right)
I_{0}\left( Q_{\min }\right) ,  \label{ap4} \\
\zeta ^{(1)}(Q,Q^{\prime }) &=&4\pi \ell ^{2}RK_{1}\left( Q_{\max }\right)
I_{1}\left( Q_{\min }\right) ,  \nonumber
\end{eqnarray}
Note, that the function $K_{0}\left( Q\right) $ diverges logarithmically at $%
Q\rightarrow 0$. This divergence is discussed in Sections II and IV.

The expressions for the harmonics $\left( \text{\ref{ap3}}\right) $ can also
be simplified for the Staras correlator, $\mu =1,$ when $C_{n}^{1}\left(
\cos \phi \right) =\sin \left[ \left( n+1\right) \phi \right] /\sin \phi $, 
\begin{eqnarray*}
\int_{0}^{2\pi }C_{m}^{1}\left( \cos \phi \right) \,d\phi &=&\left[ 0,\text{ 
}m=2k+1\text{; }2\pi ,\text{ }m=2k\right] , \\
\int_{0}^{2\pi }C_{m}^{1}\left( \cos \phi \right) \cos \phi \,d\phi &=& 
\left[ 0,\text{ }m=2k\text{; }2\pi ,\text{ }m=2k+1\right] ,
\end{eqnarray*}
and the harmonics $\left( \text{\ref{ap3}}\right) $ reduce to the rapidly
converging sums of the Bessel functions with alternating coefficients. For
all other power-law correlators with different values of $\mu $ the
integration should be performed numerically.

The last group of correlators involves power-law behavior in momentum space,
Eq. $\left( \text{\ref{ee3}}\right) $. This group includes the Lorentzian in
momentum space $\lambda =0$ that was observed in Ref.\cite{fer1} and the
exponential correlator $\left( \text{\ref{e2}}\right) ,\left( \text{\ref{ap2}%
}\right) $ at $\lambda =1/2$. In general, the angular harmonics are

\begin{eqnarray}
\zeta ^{\left( 0\right) } &=&^{\,}\frac{4\pi \ell ^{2}R^{2}}{\left[ 1+\left(
Q^{2}-Q^{\prime 2}\right) ^{2}+2\left( Q^{2}+Q^{\prime 2}\right) \right]
^{\left( 1+\lambda \right) /2}}P_{\lambda }\left( \Omega \right) ,
\label{ap5} \\
\zeta ^{\left( 1\right) } &=&\frac{4\pi \ell ^{2}R^{2}/\lambda }{\left[
1+\left( Q^{2}-Q^{\prime 2}\right) ^{2}+2\left( Q^{2}+Q^{\prime 2}\right) %
\right] ^{\left( 1+\lambda \right) /2}}P_{\lambda }^{1}\left( \Omega \right)
,  \nonumber \\
\Omega  &=&\left( 1+Q^{2}+Q^{\prime 2}\right) /\sqrt{1+\left(
Q^{2}-Q^{\prime 2}\right) ^{2}+2\left( Q^{2}+Q^{\prime 2}\right) }  \nonumber
\end{eqnarray}
where $P_{\lambda }^{n}\left( \Omega \right) $ are the associated Legendre
functions of the first kind. Note, that the argument $\Omega $ of the
Legendre functions in our expressions is larger than $1$. One should be
cautious when doing calculations with expressions $\left( \text{\ref{ap5}}%
\right) $: some of the handbooks (and software packages, {\it e.g.,
Mathematica}) do not use the same normalization for Legendre polynomials and
Legendre functions, {\it i.e., }for functions $P_{\lambda }^{n}\left( \Omega
\right) $\ with integer and non-integer $\lambda $.

In the case of the Lorentzian in momentum space, $\lambda =0,$

\begin{equation}
\zeta _{ik}\left( s\right) =\ell ^{2}K_{0}\left( s/R\right) ,\ \zeta \left(
Q\right) =\frac{2\pi \ell ^{2}R^{2}}{1+(QR)^{2}},  \label{ap6}
\end{equation}
the harmonics

\begin{eqnarray}
\zeta ^{\left( 0\right) }\left( q_{j},q_{j^{\prime }}\right) &=&^{\,}\frac{%
4\pi \ell ^{2}R^{2}}{\sqrt{1+\left( Q^{2}-Q^{\prime 2}\right) ^{2}+2\left(
Q^{2}+Q^{\prime 2}\right) }}  \label{ap7} \\
\zeta ^{\left( 1\right) }\left( q_{j},q_{j^{\prime }}\right) &=&\,\frac{8\pi
\ell ^{2}R^{2}QQ^{\prime }}{\sqrt{1+\left( Q^{2}-Q^{\prime 2}\right)
^{2}+2\left( Q^{2}+Q^{\prime 2}\right) }}  \nonumber \\
&&\times \frac{1}{1+Q^{2}+Q^{\prime 2}+\sqrt{1+\left( Q^{2}-Q^{\prime
2}\right) ^{2}+2\left( Q^{2}+Q^{\prime 2}\right) }}.  \nonumber
\end{eqnarray}
Note, that this correlator diverges in real space at $s\rightarrow 0$.


\section{Appendix B. Positions of new QSE oscillations.}

The peak positions are determined by the condition that the absolute value
of the diagonal and the first off-diagonal matrix elements in transport
equation (\ref{ee6}) become comparable: 
\[
1/\tau _{j,j+1}\sim 1/\tau _{jj} 
\]

Rewriting this condition via transition probabilities $W_{jj^{\prime
}}^{(0,1)}\left( {\bf q,q}^{\prime }\right) $ we get 
\begin{equation}
\left[ W_{jj}^{\left( 0\right) }\left( x,z\right) -W_{jj}^{\left( 1\right)
}\left( x,z\right) \right] +\sum_{j^{\prime }\neq j}W_{j,j^{\prime
}}^{\left( 0\right) }\left( x,z\right) \sim W_{j,j+1}^{\left( 1\right)
}\left( x,z\right) ,  \label{apb1}
\end{equation}
where $W_{jj^{\prime }}^{\left( 0,1\right) }\left( q_{j},q_{j^{\prime
}}\right) $ are the zeroth and first harmonics of $W\left( {\bf q}_{j}{\bf -q%
}_{j^{\prime }}\right) $\ over the angle $\widehat{{\bf q}_{j}{\bf q}}%
_{j^{\prime }}$ that can be expressed explicitly via the surface correlation
functions [see Eq. (\ref{rrr3}) and Appendix A]. For large $q_{j}R$, the
off-diagonal scattering probabilities $W_{jj^{\prime }}$ are exponentially
suppressed for Gaussian and power-law inhomogeneities, Eqs. (\ref{ap1}) ,(%
\ref{ap3}) : $W_{jj}^{\left( 0\right) }\sim W_{jj}^{\left( 1\right) }\gg
W_{j,j+1}^{\left( 0\right) }\sim W_{j,j+1}^{\left( 1\right) }$. With a
logarithmic accuracy, the condition (\ref{apb1}) corresponds to the equation 
\begin{equation}
W_{jj}^{\left( 0\right) }\left( x,z\right) -W_{jj}^{\left( 1\right) }\left(
x,z\right) =W_{j,j+1}^{\left( 0\right) }\left( x,z\right)  \label{apb2}
\end{equation}

Taking into consideration the asymptotic behavior for modified Bessel
function in Eq.(\ref{ap1}) for the Gaussian correlator, Eq.(\ref{apb2}) can
be reduced to

\begin{equation}
\frac{j^{2}}{2Q_{j}^{3}}=\frac{\left( j+1\right) ^{2}}{\sqrt{Q_{j}Q_{j+1}}}%
\exp \left[ -\frac{1}{2}\left( Q_{j}-Q_{j+1}\right) ^{2}\right] ,
\label{apb3}
\end{equation}
where $Q_{j}=x\sqrt{1-\left( \pi j/z\right) ^{2}}$. When $z/\pi j\gg 1$, we
can put $Q_{j}\approx Q_{j+1}\approx x$ in the denominator. The exponent
should be evaluated more carefully: $Q_{j}-Q_{j+1}\approx x\pi ^{2}\left(
2j+1\right) /2z^{2}.$ Then Eq. (\ref{apb3}) yields the following values of
the peak positions:

\begin{equation}
z_{j}\left( x\right) =\frac{\pi }{2}\frac{\sqrt{x\left( 2j+1\right) }}{\left[
\ln \left( x\sqrt{2}\frac{1+j}{j}\right) \right] ^{1/4}}.  \label{apb4}
\end{equation}
Since $x=yz$, these peak positions $z_{j}\left( x\right) $\ can be also used
to get the peak positions for the conductivity at fixed $y$, $z_{j}\left(
y\right) $ as a solution of the following algebraic equation: 
\begin{equation}
z_{j}\left( y\right) =\frac{\pi ^{2}}{4}\frac{y\left( 2j+1\right) }{\left[
\ln \left( z_{j}\left( y\right) y\sqrt{2}\frac{1+j}{j}\right) \right] ^{1/2}}%
.  \label{apb5}
\end{equation}

Similar but more cumbersome calculations, can be performed for the power-law
correlators (\ref{ee2}) . For example, if $\mu =1/2$, Eq. (\ref{apb2}) reads 
\begin{equation}
4j^{2}\int\limits_{0}^{\pi /2}\exp \left( -2Q_{j}\sin t\right) \,\sin
^{2}t\,dt=2\left( j+1\right) ^{2}\int\limits_{0}^{\pi /2}\exp \left( -\sqrt{%
\nu ^{2}+4Q_{j}Q_{j+1}\sin ^{2}t}\right) \,dt,  \label{apb6}
\end{equation}
where we introduced $\nu \equiv \nu _{j,j+1}=Q_{j}-Q_{j+1.\text{ }}$For
large $Q_{j},$ an asymptotic estimate for the integral in the left-side is $%
1/4Q_{j}^{3}$. Rough asymptotic estimate for the integral in the right-side
of the equation is 
\[
\int\limits_{0}^{1}\exp \left( -\sqrt{\nu ^{2}+4Q_{j}Q_{j+1}\,t^{2}}\right)
\,\frac{dt}{\sqrt{1-t^{2}}}\approx \frac{1}{2\sqrt{Q_{j}Q_{j+1}}}%
\int\limits_{0}^{2\sqrt{Q_{j}Q_{j+1}}}\exp \left( -\sqrt{\nu ^{2}+y^{2}}%
\right) \,dy 
\]
In order to estimate this integral, we can substitute $\sqrt{\nu ^{2}+y^{2}}$
by 
\[
\sqrt{\nu ^{2}+y^{2}}\rightarrow \left\{ 
\begin{array}{cc}
\nu , & \text{for }y<\nu \\ 
y, & \text{for }y>\nu .
\end{array}
\right. 
\]
Then 
\[
\frac{1}{2Q_{j}}\int\limits_{0}^{\infty }\exp \left( -\sqrt{\nu ^{2}+y^{2}}%
\right) \,dy\approx \frac{1}{2Q_{j}}e^{-\nu }\left( \nu +1\right) . 
\]
This leads to the following estimate for peak positions 
\begin{equation}
z_{j}\left( x\right) =\pi \sqrt{\frac{x\left( 2j+1\right) }{2\nu _{j}}},
\label{apb7}
\end{equation}
where $\nu _{j}$ is the root of the transcendental equation 
\[
\nu _{j}=2\ln A_{j}+\ln \left( 1+\nu _{j}\right) ,,\ A_{j}\equiv x\left(
1+1/j\right) . 
\]
The last equation can be solved by iterations: 
\begin{eqnarray*}
\nu &=&\nu ^{\left( 0\right) }+\nu ^{\left( 1\right) }+\nu ^{\left( 2\right)
}+..., \\
\nu ^{\left( 0\right) } &=&2\ln A,\;\nu ^{\left( 1\right) }=\ln \left[ 2\ln
A+1\right] ,...\;
\end{eqnarray*}

Finally, with a logarithmic accuracy, the solution of Eq. (\ref{apb6}) for
the positions of peaks becomes 
\begin{equation}
z_{j}=\frac{\pi }{2}\sqrt{\frac{x\left( 2j+1\right) }{\ln \left[ x\sqrt{\ln
\left( x(1+1/j\right) }\left( 1+1/j\right) \right] }}.  \label{apb8}
\end{equation}
Similar asymptotic estimates for the power-law correlators with arbitrary $%
\mu $ yield

\begin{eqnarray}
z_{j} &=&\frac{\pi }{2}\sqrt{\frac{x\left( 2j+1\right) }{\ln \left[
A_{j}\left( 2\ln A_{j}\right) ^{\mu /2+1/4}\right] }},  \nonumber \\
A_{j} &=&x\frac{1+j}{j}\sqrt{\frac{2}{\Gamma \left( \mu +5/2\right) }}.
\label{apb9}
\end{eqnarray}
It is clear from Eq. (\ref{apb9}) that the dependence of the peak positions
on $\mu $ is extremely weak. 


\end{document}